\documentclass{aa}
\usepackage{amsmath}
\usepackage{amssymb}
\usepackage[varg]{txfonts}
\usepackage{epsfig,graphicx}
\usepackage[colorlinks=true, linkcolor=blue, citecolor=blue, urlcolor=blue]{hyperref}  
\usepackage{natbib}
\bibpunct{(}{)}{;}{a}{}{,} 
\newcommand{\ser}{S\'ersic}
\usepackage{color}

\defcitealias{2015MNRAS.451.1728D}{DG15}
\defcitealias{1999A&A.344.868X}{X99}
\begin{document}

\title{\emph{HERschel} Observations of Edge-on Spirals (\textit{HER}OES)}
\subtitle{III. Dust energy balance study of IC\,2531\thanks{{\it Herschel} is an ESA space observatory with science instruments provided by European-led Principal Investigator consortia and with important participation from NASA.} }

\titlerunning{Dust energy balance study of the {\it HER}OES galaxy IC\,2531}

\author{
Aleksandr~V.~Mosenkov\inst{1,2,3}
\and Flor~Allaert\inst{1}
\and Maarten~Baes\inst{1}
\and Simone~Bianchi\inst{4}
\and Peter~Camps\inst{1}
\and Gert~De Geyter\inst{1}
\and Ilse~De Looze\inst{5,1}
\and Jacopo~Fritz\inst{6}
\and Gianfranco~Gentile\inst{1,7}
\and Thomas~M.~Hughes\inst{8}
\and Fraser~Lewis\inst{9,10}
\and Joris~Verstappen\inst{11}
\and Sam~Verstocken\inst{1}
\and S\'ebastien~Viaene\inst{1}}
\institute{Sterrenkundig Observatorium, Universiteit Gent, Krijgslaan 281, B-9000 Gent, Belgium\\  
\email{Aleksandr.Mosenkov@UGent.be}
\and
St.Petersburg State University, Universitetskij pr. 28, 198504, St. Petersburg, Stary Peterhof, Russia
\and
Central Astronomical Observatory of RAS, Pulkovskoye chaussee 65/1, 196140, St. Petersburg, Russia
\and Dipartimento di Matematica e Fisica, Universit\`{a} degli Studi Roma Tre, Via della Vasca Navale 84, 00146, Roma, Italy
\and Department of Physics and Astronomy, University College London, Gower Street, London WC1E 6BT, UK
\and Centro de Radioastronom\'{i}a y Astrof\'{i}sica, CRyA, UNAM, Campus Morelia, A.P. 3-72, C.P. 58089, Michoac\'{a}n, Mexico 
\and Department of Physics and Astrophysics, Vrije Universiteit Brussel, Pleinlaan 2, 1050 Brussels, Belgium
\and Instituto de F\'{i}sica y Astronom\'{i}a, Universidad de Valpara\'{i}so, Avda. Gran Breta\~{n}a 1111, Valpara\'{i}so, Chile
\and Faulkes Telescope Project, Cardiff University, The Parade, Cardiff CF24 3AA, Cardiff, Wales
\and Astrophysics Research Institute, Liverpool John Moores University, IC2, Liverpool Science Park, 146 Brownlow Hill, Liverpool L3 5RF, UK
\and Kapteyn Astronomical Institute, University of Groningen, Landleven 12, Groningen NL-9747AD, The Netherlands
}
\date{}
\abstract{We investigate the dust energy balance for the edge-on galaxy IC\,2531, one of the seven galaxies in the {\it HER}OES sample. We perform a state-of-the-art radiative transfer modelling based, for the first time, on a set of optical and near-infrared galaxy images. We show that taking into account near-infrared imaging in the modelling significantly improves the constraints on the retrieved parameters of the dust content. We confirm the result from previous studies that including a young stellar population in the modelling is important for explaining the observed stellar energy distribution. However, the discrepancy between the observed and modelled thermal emission at far-infrared wavelengths, the so-called dust energy balance problem, is still present: the model underestimates the observed fluxes by a factor of about two. We compare two different dust models, and find that dust parameters and thus the spectral energy distribution in the infrared domain are sensitive to the adopted dust model. In general, the THEMIS model reproduces the observed emission in the infrared wavelength domain better than the popular Zubko et al.\ BARE-GR-S model. Our study of IC\,2531 is a pilot case for detailed and uniform radiative transfer modelling of the entire {\it HER}OES sample, which will shed more light on the strength and origins of the dust energy balance problem.}


\maketitle

\section{Introduction}

Cosmic dust is one of the fundamental components of the interstellar medium (ISM). Notwithstanding its small contribution to the total mass of a galaxy, interstellar dust plays an important role in many physical and chemical processes in galaxies. It is well established that dust catalyses the transformation of atomic to molecular hydrogen \citep[see e.g.][]{2007ApJ.654.273G} and contributes to the cooling and heating of the ISM \citep{Tielens2005}. Moreover, dust efficiently absorbs and scatters light at UV and optical wavelengths, and re-emits the absorbed energy as thermal emission at far-infrared (FIR) and submillimeter wavelengths \citep{1991AJ.101.354S}. On average, about one third of all starlight in normal late-type galaxies is absorbed by dust \citep{2002MNRAS.335L..41P, 2016A&A...586A..13V}. A solid knowledge of the amount of dust, its physical properties, its spatial distribution and the heating mechanisms of the interstellar medium are hence crucial for modern astrophysics.

According to the dust energy balance, we would expect the absorbed energy to match the re-emitted energy in the IR/submm domain. However, an inconsistency is generally found: the FIR emission predicted from model based on extinction at optical wavelengths typically underestimates the observed values by a factor of 3--4. This is the well-known dust energy balance problem that has been observed in many studies \citep{2000A&A...359...65B, 2000A&A...362..138P, 2011A&A...527A.109P, 2001A&A...372..775M, 2004A&A...425..109A, 2005A&A...437..447D, 2008A&A...490..461B, 2010A&A...518L..39B, 2012A&A...541L...5H, 2015MNRAS.451.1728D}.

Edge-on galaxies are of particular interest as these objects offer a unique perspective to study the vertical as well as the radial structure of their stellar components. Moreover, as interstellar dust obscures a large fraction of the emitted starlight, they are ideal targets to study the dusty ISM in spiral galaxies. Both the stellar structure and dust distribution in edge-on galaxies have been considered in multiple studies since the early 1980s \citep{vderk1981a, 1994A&AS..103..475B, 1999A&A.344.868X, pohlen+2000, kregel+2002, 2007A&A...471..765B, 2010A&A...518L..39B, 2010MNRAS.401.559M}. In particular, the prototypical edge-on spiral galaxy, NGC\,891, has been studied extensively over the past three decades \citep[e.g.][]{vderk1981b, 1998A&A.331.894X, 2000A&A...362..138P, 2011A&A...527A.109P, 2011A&A.531L.11B, 2012ApJ...746...70S, 2014A&A...565A...4H, 2015A&A.575A.17H}. The most suitable approach to derive the intrinsic properties of stars and dust is through dust radiative transfer modelling \citep[for a review, see][]{2013ARA&A..51...63S}. Several radiative transfer codes now incorporate a full treatment of absorption, multiple scattering and thermal emission by dust \citep[e.g.,][]{2001ApJ...551..269G, 2003MNRAS.343.1081B, 2006MNRAS.372....2J, 2008A&A...490..461B, 2011A&A...536A..79R}.

Recently, \citet{2015MNRAS.451.1728D} studied the dust energy balance in two edge-on galaxies IC\,4225 and NGC\,5166 observed by Herschel in the frame of the Herschel-ATLAS \citep{2010PASP..122..499E}. They fitted detailed radiative transfer models directly to a set of optical images, and included an additional young stellar disc to match the UV fluxes. They concluded that this additional component is necessary in order to reproduce the heating of the dust grains. For NGC\,5166, the predicted spectral energy distribution (SED) and the model images match the observations very well. However, for IC\,4225 the far-infrared emission of their radiative transfer model still underestimates the observed fluxes by a factor of three pointing to the same dust energy problem. One of the possible reasons of that discrepancy could be that the angular size of IC\,4225 is too small to properly fit the dust disc parameters. Among the other reasons can be emission of obscured star-forming regions deeply embedded in dense dust clouds which do not contribute notably to the observed UV flux but have a clear impact on the FIR emission \citep[see e.g.][]{2012MNRAS.419..895D, 2014A&A...571A..69D, 2015MNRAS.451.1728D}.

In the present work, we investigate the dust energy balance in the edge-on galaxy IC\,2531. It is one of the galaxies from the {\it HER}OES \citep{2013A&A...556A..54V} sample, a set of seven edge-on spiral galaxies. All of these galaxies have optical diameter of at least 4 arcmin, so that they are well-resolved, even at infrared and submm wavelengths. They were selected to have a clear and regular dust lane, and they have already been fitted with a radiative transfer code based on their optical and NIR data. All galaxies were observed with the {\it Herschel Space Observatory} \citep{2010A&A...518L...1P}, in five PACS \citep{2010A&A...518L...2P} and SPIRE \citep{2010A&A...518L...3G} bands at 100, 160, 250, 350 and 500~$\mu$m. The general goal of the {\it HER}OES project is to present a detailed, systematic and homogeneous study of the amount, spatial distribution and properties of the interstellar dust in these seven galaxies, and investigate the link between the dust component and stellar, gas and dark matter components. The first paper in this series \citep{2013A&A...556A..54V} presented the \textit{Herschel} imaging data, and the second one \citep{2015A&A...582A..18A} presented H{\sc{i}} observations and detailed tilted ring models. The main goals of this third paper is to study the dust energy balance in IC\,2531, and to develop an approach that will be used for studying the dust in all {\it HER}OES galaxies in a uniform and consistent way.   

The outline of this paper is as follows. In Section~\ref{sec:target}, we describe the main properties of IC\,2531 taken from the literature. In Section~\ref{sec:data}, we present the available imaging and describe the data reduction. In the following two sections we describe our modelling approach and results (our approach is based on \citet{2015MNRAS.451.1728D}, but we use a wider range of imaging data and high-resolution imaging, which forces us to adapt the modelling strategy). In Section~\ref{sec:IRACdecomp}, we present a detailed photometrical decomposition of the IRAC 3.6~$\mu$m image into several stellar components. In Section~\ref{sec:fitskirt}, we  perform an oligochromatic radiative transfer fitting for this galaxy using seven optical and NIR bands in order to constrain the main dust disc parameters. The panchromatic radiative transfer modelling of the galaxy based on the retrieved dust and stellar component models is presented in Section~\ref{sec:simulations}, where we compare different dust mixture models as well. Finally, we discuss our findings in Section~\ref{sec:discussion} and present a summary in Section~{\ref{sec:summary}}.

\section{The target} 
\label{sec:target}

\begin{figure*}
\includegraphics[width=\textwidth , angle=0, clip=]{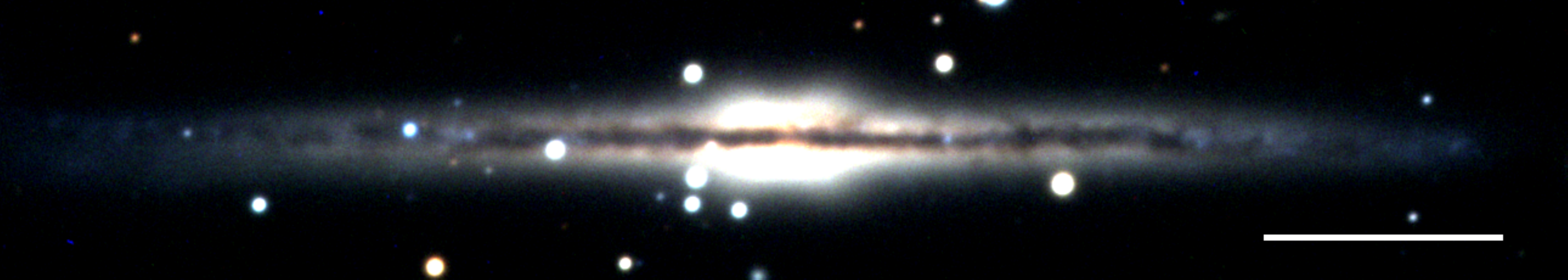}
\caption{Composite RGB-image of the $B$-, $V$- and $R$-passband frames from the Faulkes Telescope South (see Sect.~\ref{sec:data}). The length of the white bar in the bottom right corner is 1~arcmin.}
\label{IC2531_rgb}
\end{figure*}

IC\,2531 is a nearby late-type spiral galaxy ($D\approx36.8$~Mpc; see Table~\ref{table1}) located in the southern hemisphere. It is viewed almost directly edge-on: in Fig.~\ref{IC2531_rgb} we can see that the sharp dust lane divides the galaxy body into two equal parts. The fitted inclination of $89\fdg6\pm0\fdg2$ from \citet{1999A&A.344.868X} confirms that it is very close to exactly edge-on. The boxy/peanut-shaped bulge is a prominent detail of this galaxy and has been studied in several works \citep[e.g.][]{2006MNRAS.366.1121P, 2006MNRAS.370..753B}, and also a slight warp is visible at optical wavelengths. Despite of these peculiarities in the galaxy structure, we choose IC\,2531 as a pilot case, because it has an overall regular structure and is sufficiently large that the observations in the whole UV-submm domain can be analysed.

A thorough analysis of the H{\sc{i}} content of IC\,2531 was carried out by \cite{2015A&A...582A..18A} as part of the second {\it HER}OES paper \citep[see also][]{2004MNRAS.352..768K, 2010A&A...515A..60O, 2010A&A...515A..62O}. They constructed 3D tilted-ring models of the atomic gas discs of the {\it HER}OES galaxies to constrain their surface density distribution, rotation curve and geometry. For IC\,2531, a total H{\sc{i}} mass of $1.37 \times 10^{10}~$M$_\odot$ was measured. Like the stellar disc, the H{\sc{i}} disc was found to be more extended towards the NE (approaching) side. This side also shows significantly more irregular kinematical behaviour in the H{\sc{i}} data cube than the receding (SW) side. The outer regions of the disc were found to be slightly warped both in the plane of the sky and along the line of sight, with the position and inclination angles deviating by up to about 4 degrees with respect to the centre of the galaxy. The best fitting model uses a constant scale height $z_{0}$ of 1.0 kpc. Within the errors on $z_{0}$, however, a modest flaring of the gas disc is also possible. Finally, a bright ridge in the major axis H{\sc{i}} position-velocity diagram suggests the presence of a prominent spiral arm in the disc.

\begin{table}
\centering
\caption{Basic properties of IC\,2531.}
\label{table1}
\begin{tabular}{cc}
\hline\\[-1ex]
RA (J2000)  & 09:59:56 \\
Dec (J2000) & -29:37:01 \\
Type        & Sc \\
$m_\mathrm{V}$ (mag) & 12.0 \\
Semi-major axis (arcmin) & 3.5 \\
Semi-minor axis (arcmin) & 0.4 \\
$D$ (Mpc)   & 36.8 \\
Scale (pc/arcsec) & 178 \\
$M_\mathrm{V}$ (mag)& -20.8 \\
$i$ (deg)         & 89.6 \\[0.5ex]
\hline \\
\end{tabular}
\parbox[t]{90mm}{ {\bf Notes.} The celestial coordinates and morphology are taken from NASA/IPAC Extragalactic Database (NED). The distance $D$ with the corresponding scale is taken from \cite{2013A&A...556A..54V} which was found as the average value of the redshift independent distance measurements (mostly based on Tully-Fisher relation). The apparent magnitude $m_\mathrm{V}$, semi-major and semi-minor axes are measured for the isophote of 25 mag/arcsec$^2$. The absolute magnitude $M_\mathrm{V}$ corresponds to the adopted distance $D$ and apparent magnitude $m_\mathrm{V}$, corrected for Galactic extinction according to \cite{2011ApJ...737..103S}. The inclination of the galaxy was derived from radiative transfer modelling (see Sect.~\ref{sec:fitskirt}).}
\end{table} 

\begin{table*}
\centering
\caption{Structural parameters of the stellar and dust components from different studies.}
\label{Disc_pars_authors}
\begin{tabular}{ccccccccc}
\hline
\hline\\[-2ex]
Reference & Band & $i$ & $h_\mathrm{R,*}$ & $h_\mathrm{z,*}$ & $h_\mathrm{R,d}$ & $h_\mathrm{z,d}$ & $r_\mathrm{e,b}$ & $\tau_\lambda^f$ \\ 
          &      &(deg)&      (kpc)       &        (kpc)     &      (kpc)       &       (kpc)      &      (kpc)       &  \\
\hline\\[-1ex]
(1) & $B$ & $89.6\pm0.2$ & $11.34\pm0.17$ & $0.72\pm0.02$ & $14.85\pm0.50$ & $0.45\pm0.02$ & $3.28\pm0.25$ & $0.40\pm0.01$\\
(1) & $V$ & $89.6\pm0.2$ & $8.73\pm0.17$ & $0.69\pm0.02$ & $13.68\pm0.33$ & $0.38\pm0.02$ & $2.06\pm0.13$ & $0.30\pm0.01$\\
(1) & $I$ & $89.6\pm0.2$ & $8.45\pm0.17$ & $0.72\pm0.02$ & $14.10\pm0.33$ & $0.35\pm0.02$ & $2.63\pm0.13$ & $0.22\pm0.01$\\
(1) & $J$ & $89.7\pm0.2$ & $8.30\pm0.50$ & $0.74\pm0.17$ & $13.51\pm1.00$ & $0.33\pm0.12$ & $3.29\pm0.17$ & $0.06\pm0.02$\\
(1) & $K$ & $89.6\pm0.2$ & $8.43\pm0.17$ & $0.75\pm0.03$ & $13.38\pm0.50$ & $0.37\pm0.05$ & $3.34\pm0.40$ & $0.02\pm0.01$\\
(2) & $I$   &  --- & $17.3\pm0.9$ & $0.9\pm0.1$ & --- & --- & $1.79\pm0.05$ & --- \\      
(3) & $K_s$ &  --- & $6.36\pm0.57$ & $0.46$ & --- & --- & --- & --- \\[0.5ex]  

\hline\\
\end{tabular}  
\parbox[t]{180mm}{ {\bf Notes.} (1) \cite{1999A&A.344.868X}, (2) \cite{kregel+2002}, (3) \cite{2009ApJ...702.1567B}.  The sizes have been rescaled to the distance presented in Table~\ref{table1}. The models from (2) and (3) assume a perfect edge-on disc orientation.}
\end{table*} 


The structural parameters of the stellar and dust disc of IC~2531 have been already estimated by several authors (see Table~\ref{Disc_pars_authors}). The most advanced modelling was undertaken by \citet{1999A&A.344.868X}, who fitted a radiative transfer model to several optical and near-infrared images. They adopted a model consisting of a double-exponential distribution for the dust and stellar disc plus a de Vaucouleurs profile for the bulge.

The two other works referenced in Table~\ref{Disc_pars_authors} assume a perfectly edge-on orientation of the galaxy. In \cite{kregel+2002}, a simple 2D decomposition into a de Vaucouleurs bulge and the double-exponential disc was performed. It is interesting to note that the stellar disc scale length appeared to be twice as large as was found in \citet{1999A&A.344.868X}. On the other hand, the bulge is significantly more compact as compared to the same work. These differences of the stellar components may be caused by different fitting techniques or different quality of the data used in their analysis. \cite{2009ApJ...702.1567B} determined both the radial and vertical scale parameters fitting a number of photometrical cuts parallel to the minor axis of the galaxy. Based on the analysis of 2MASS images, they found a model of the stellar disc which is 1.5--2 times thinner than in the previous studies. 



A direct comparison of the results from Table~\ref{Disc_pars_authors} is difficult, even at a fixed wavelength, since the different studies used techniques, with different levels of sophistication. As demonstrated by \cite{2015MNRAS.451.2376M}, this can seriously affect the resulting galaxy parameters. Particularly striking is the huge discrepancy (more than a factor two) between the I-band radial scale length obtained by \cite{kregel+2002} compared to the one obtained by \citet{1999A&A.344.868X}. The former model seems to be somewhat too simplistic since it does not take into account disc truncation and the attenuation by dust. The latter model, on the other hand, is a full-scale radiative transfer model that properly takes into account absorption and scattering by interstellar dust.  

All in all, the above mentioned studies show that IC~2531 has a more than average radially extended stellar disc (radial scale $h_\mathrm{R,*}\ga6-8$~kpc) with a stronger than average disc flattening ($h_\mathrm{R,*}/h_\mathrm{z,*}\ga13$) and an extended dust component.  

\section{Observations and data preparation} 
\label{sec:data}

Optical observations of IC\,2531 were made at the {\it Faulkes Telescope South} located at Siding Spring, New South Wales, Australia. The galaxy was observed in the $B$-, $V$- and $R$-band filters with the 2-m telescope in February 2014. For each band, nine individual images with exposure times of 200 s were cleaned, aligned and combined into a single science frame using standard data reduction techniques, including bias subtraction, flat-fielding and removing instrumental artefacts. The final images have a pixel scale of 0.3\arcsec /pix and an average seeing of 1.5\arcsec. A three-colour image combined from all three bands is presented in Fig~\ref{IC2531_rgb}. 

Additional near-infared imaging in the $J$, $H$ and $K_\mathrm{s}$ passbands was retrieved from the All-Sky Data Release of the {\it Two Micron All Sky Survey} (2MASS, \citealp{2006AJ....131.1163S}). Although the resolution of these images is rather poor, the angular size of the galaxy along with the signal-to-noise ratio is sufficient for our further analysis.

Furthermore, we use mid-IR data from the {\it Spitzer Space Telescope} \citep{2004ApJS..154....1W}. In this paper we will make use of the Infrared Array Camera (IRAC, \citealp{2004ApJS..154...10F}), namely 3.6 $\mu\mathrm{m}$ observations available through the {\it Spitzer} Heritage Archive (the mosaic {\it *maic.fits} and uncertainties {\it *munc.fits} files from post-Basic Calibrated Data).

Before the imaging data can be analysed with our methods (see Sect.~\ref{sec:IRACdecomp} and Sect.~\ref{sec:fitskirt}), some preparation of the retrieved images is needed. This includes astrometry correction in the same way for all frames, rebinning (i.e. resampling, such that all the data would have the same pixel scale), subtraction of the sky background, rotating the images (the galaxy should be aligned with the horizontal axis), cropping the frames to the appropriate size to cover the whole galaxy with very little empty space, and, finally, masking all foreground and background objects contaminating the galaxy image. In addition, the point spread function (PSF) was carefully determined for all the frames. 

To verify astrometry, we used the astrometric calibration service Astrometry.net\footnote{http://nova.astrometry.net/}. For the next steps we applied the \textsc{python} Toolkit for \textsc{skirt}\footnote{The latest version is available at http://www.skirt.ugent.be/pts/} (\textsc{pts}) which includes special routines to prepare data for the consequent analysis with the \textsc{fitskirt} code. The average alignment of the rebinned images is estimated to be 0.5 pix. Near-infrared observations are strongly affected by thermal radiation from the sky and the telescope, hence, the sky background should be estimated very accurately. To do so, we inspected each frame separately and selected empty parts (using the free software SAOimage DS9\footnote{http://ds9.si.edu/site/Home.html}) in each frame around the outermost isophotes of the galaxy. The intensities in selected regions were then fitted with a $2^\mathrm{nd}$ order polynomial and the fitted sky frames were then subtracted from the respective images. Subsequently, the galaxy images were rotated \citep[the positional angle was measured for the IRAC\,3.6 $\mu\mathrm{m}$ image in the way proposed in][sect.~3.1.2]{2012MNRAS.427.1102M}, cut out from the initial frames, such that the galaxy 25.5 mag/arcsec$^2$ isophote in the $B$-band would be fully included. The final images were thoroughly revisited by hand to mask stars, background galaxies and other contaminants. Finally, we obtained the prepared images of the galaxy in seven bands ready for the fitting analysis. We will call them the reference images as our further modelling will be directly based on them.

Apart from the imaging described above, we collected imaging data at other wavelengths, from the far-ultraviolet to the submm. These observations will be taken into account during the panchromatic modelling described in Sect.~\ref{sec:simulations}. {\it Galaxy Evolution Explorer} ({\it GALEX}, \citealp{2005ApJ...619L...1M, 2014AdSpR..53..900B}) observations at 0.152\,$\mu\mathrm{m}$ (FUV) and 0.227\,$\mu\mathrm{m}$ (NUV) were downloaded from the All Sky Imaging Survey reachable trough the GalexView service\footnote{http://galex.stsci.edu/GalexView}. IC\,2531 was also observed by the {\it Wide-field Infrared Survey Explorer} ({\it WISE}, \citealp{2010AJ....140.1868W}), and we downloaded the frames of the galaxy in all four passbands. {\it{Herschel}} imaging was taken from \cite{2013A&A...556A..54V}, but the reduction was repeated using the latest calibration pipeline.  

All the images were prepared in exactly the same way as the reference images, with the same rebinning and alignment. Integrated flux densities for all wavelengths were determined by means of aperture photometry from the \textsc{photutils} \textsc{python} package. We used the same mask for all galaxy images, and the masked areas were filled with the interpolated values using a Gaussian weights kernel. Finally, we added flux densities (APERFLUX) at 353, 545 and 857~GHz (850\,$\mu\mathrm{m}$ , 550\,$\mu\mathrm{m}$ and 350\,$\mu\mathrm{m}$, respectively) from the {\it Planck} Legacy Archive\footnote{http://pla.esac.esa.int/pla} \citep{2014A&A...571A...1P}.  The final global fluxes are listed in Table~\ref{tab:Fluxes.tab}.

\begin{table}
\caption{Observed flux densities and their corresponding errors. Correction for Galactic extinction (where applicable) has been applied according to \cite{2011ApJ...737..103S}. The masked objects were replaced by the interpolated values using a Gaussian weights kernel.}
\label{tab:Fluxes.tab}
\centering
\begin{tabular}{ccc}
\hline
\hline\\[-2ex]
Survey & $\lambda$ & $F_\nu$ \\[-0.5ex]
       & ($\mu\mathrm{m}$) & (mJy) \\[0.5ex]
\hline\\[-1ex]
{\it GALEX}   & 0.153 & $1.02\pm0.48$ \\
{\it GALEX}   & 0.226 & $1.90\pm0.35$ \\
FTS     & 0.45  & $40.27\pm4.3$ \\
FTS     & 0.55  & $70.22\pm6.6$ \\
FTS     & 0.66  & $97.6\pm8.8$ \\
2MASS   & 1.25  & $250.0\pm16.4$ \\
2MASS   & 1.64  & $346.8\pm19.5$ \\
2MASS   & 2.17  & $290.9\pm14.5$ \\
{\it WISE}    & 3.4   & $127.3\pm5.4$ \\
{\it Spitzer} & 3.6   & $156.3\pm7.6$ \\
{\it WISE}    & 4.6   & $88.2\pm3.7$ \\
{\it WISE}    & 12    & $218.2\pm6.8$ \\
{\it WISE}    & 22    & $187.9\pm7.7$ \\
PACS          & 100   & $5824\pm508$ \\
PACS          & 160   & $9937\pm654$ \\
SPIRE         & 250   & $6720\pm477$ \\
SPIRE         & 350   & $3511\pm256$ \\
{\it Planck}        & 350   & $3499\pm552$ \\
SPIRE         & 500   & $1419\pm115$ \\
{\it Planck}        & 550   & $1180\pm290$ \\
{\it Planck}        & 850   & $313\pm174$ \\

\hline
 \end{tabular}
\end{table}

\section{Oligochromatic radiative transfer modelling}

The first step in our modelling consists of the construction of a radiative transfer model that can reproduce the images at optical (and near-infrared) wavelengths. The most straightforward approach would be to follow the strategy of \citet{2014MNRAS.441..869D}. They studied a sample of 12 edge-on spiral galaxies, and directly fitted a Monte Carlo radiative transfer model to the SDSS images. Their three-component model consisted of a double-exponential disc to describe the stellar and dust discs and a \ser\ profile to describe the bulge. Though their model contained 19 free parameters, they managed to accurately reproduce almost all the galaxies in the sample, without human intervention or strong boundary conditions. 

In this work, we use a different approach. The main driver for this is that, since the {\it{HER}}OES galaxies are so nearby, the images have higher resolution and this three-component model is too simplified. It is well known that the Milky Way has a thin and a thick disc \citep{1983MNRAS.202.1025G}. Many edge-on galaxies studied within the frame of the S$^4$G project exhibited a clear presence of a thick disc as well \citep{2012ApJ...759...98C, 2015ApJS..219....4S}. We will show later that IC\,2531 can be described more accurately if a model includes a thin and a thick disc rather than one stellar disc. Unfortunately, a model consisting of three stellar components and a dust disc would comprise too many parameters to be fitted directly in a radiative transfer fitting procedure. 

Therefore, we decided to use a two-step approach instead of the direct fitting as in \citet{2014MNRAS.441..869D}. First, we use the IRAC 3.6\,$\mu\mathrm{m}$ image of IC\,2531, for which we can neglect the obscuring effects of dust, to build a geometrical model of the stellar components. Subsequently, we will perform an oligochromatic radiative transfer fitting of optical and near-infrared images with the fixed geometrical parameters of the stellar components (but free luminosity values) and find dust disc parameters. A similar approach was applied to the elliptical galaxy NGC~4370 \citep{2015A&A...579A.103V}.  

\subsection{Decomposition of the IRAC 3.6\,$\mu\mathrm{m}$ image} 
\label{sec:IRACdecomp}

The 3.6\,$\mu\mathrm{m}$ passband offers a clear view of the old stars of a galaxy, which make up the bulk of the stellar mass. It is not strongly affected by dust extinction and is not sensitive to star forming regions, making it the best suited band for deriving the structural parameters of the stellar components. Although the resolution of the IRAC 3.6\,$\mu\mathrm{m}$-band image is worse than for the optical images, IC\,2531 is still sufficiently resolved in the {\it Spitzer} image, giving an opportunity to trace both the thin and thick disc of the galaxy (see below). 

The profile of the bulge of a spiral galaxy is usually fitted by means of a \ser\ model \citep{ser1968}:
\begin{equation}
I(r) = I_\mathrm{e,b} \: \exp \left\{ -b_n \left[ \left( \frac{r}{r_\mathrm{e,b}} \right)^{1/n_\mathrm{b}} \! - \: 1 \right] \right\},
\end{equation}
where $I_\mathrm{e,b}$ is the effective surface brightness, i.e.\ the surface brightness at the half-light radius of the bulge $r_\mathrm{e,b}$, and $b_n$ is a function of the \ser\ index $n_\mathrm{b}$ \citep{caon+1993, ciotti99}. The apparent axis ratio of the bulge $q_\mathrm{b}$ is also a free parameter of the model. The 3D luminosity density corresponding to this intensity distribution on the sky can be expressed using complex special functions \citep{2002A&A...383..384M, 2011A&A...525A.136B, 2011A&A...534A..69B}.

In order to describe the surface brightness distribution of the three-dimensional axisymmetric stellar disc, radiative transfer modellers usually use a 3D luminosity density model where the radial and vertical profile of the luminosity density are exponential (the double-exponential disc).  In a cylindrical coordinate system $(R, z)$ aligned with the disc (where the disc mid-plane has $z = 0$), the luminosity density $j(R,z)$ is given by
\begin{equation}
j(R,z) = j_{0} \; \mathrm{e}^{-R/h_R-|z|/h_z}\,,
\label{exp_disc}
\end{equation}
where $j_{0}$ is the central luminosity density of the disc, $h_R$ is the disc scale length and $h_z$ is the vertical scale height. However, analysis of surface brightness profiles of galaxies has shown that not all of galactic discs can be described accurately by Eq.~(\ref{exp_disc}), as they typically show truncations and anti-truncations \citep{erwin05, erwin08, munoz-mateos13}. A modified profile can be presented as \citep{erwin08, 2015ApJ...799..226E}: 
\begin{equation}
j(R,z) = S \, j_{0} \, 
\text{e}^{-\frac{R}{h_{R,\text{inn}}}-\frac{|z|}{h_z}} 
\left(1 + \text{e}^{\frac{s\,(R-R_\text{b})}{h_{R,\text{out}}}}\right)^{ \frac{1}{s}
\left(\frac{h_{R,\text{out}}}{h_{R,\text{inn}}} - 1\right)
}\,.
\label{eq:broken_disc}
\end{equation}
In this formula, $s$ parametrises the sharpness of the transition between the inner and outer profiles with the break radius $R_\text{b}$. Notice here that the sharpness $s$ is dimensionless, whereas the sharpness parameter $\alpha=s/h_\mathrm{R,out}$ in \cite{erwin08} is has units of length$^{-1}$. Large values of sharpness parameter ($s \gg 1$) correspond to a sharp transition, and small values ($s \sim 1$) set a very gradual break. The dimensionless quantity $S$ is a scaling factor, given by
\begin{equation}
S = \left(1 + \text{e}^{-\frac{s \, R_\text{b}}{h_{R,\text{out}}}}\right)^{-\frac{1}{s} \left(\frac{h_{R,\text{out}}}{h_{R,\text{inn}}} - 1\right)}\,.
\label{eq:broken_disc_S}
\end{equation}
Thus, this model contains five free parameters: the scale length of inner disc $h_{R,\text{inn}}$, the scale length of outer disc $h_{R,\text{out}}$, the scale height $h_z$, the break radius $R_\text{b}$, and the sharpness of the break $s$.


Visual inspection of the minor-axis profile from the IRAC 3.6~$\mu$m image for IC\,2531 (see Fig.~\ref{IC2531_irac_profiles},  left plot) suggests the presence of both a thin and a thick disc (as an excess over a simple exponential thin disc at the distance $|z|\ga15\arcsec$). Also, a slight change of the disc profile slope is visible on the horizontal distribution of the surface brightness at the radius $|r|\approx120\arcsec$ which is evidence for the presence of breaks in one or both stellar discs. Therefore, we performed decomposition of the 3.6~$\mu$m-band galaxy image into three stellar components: thin and thick discs with breaks plus a central \ser\ component for the bulge. 

For this purpose, we used the \textsc{imfit} code which works with the 3D disc and has the {\it BrokenExponentialDisk3D} function identical to the model (\ref{eq:broken_disc}) (with $n=100$ in Eqs.~(40) and (41) in \cite{2015ApJ...799..226E} to describe the exponential disc model). For PSF convolution, we used an in-flight point response function (PRF) image for the center of the IRAC 3.6\,$\mu\mathrm{m}$ field,\footnote{http://irsa.ipac.caltech.edu/data/SPITZER/docs/irac/calibrationfiles} downsampled to the 0.6\arcsec{} pixel scale and re-rotated to correspond to the analysed galaxy frame. Taking into account that the uncertainty of the galaxy inclination in \citet{1999A&A.344.868X} is small ($0\fdg2$), the disc inclinations in our model were set to be the same for both discs and fixed at $89\fdg6$ to minimise the number of free parameters. A fixed sharpness of the breaks, $s=5$, was applied for both discs as well. We tried using different values of $s$ (in the range from 0.1 to 100) and found that this parameter does not affect the results of the fitting for this galaxy. We used a genetic algorithm (which is also available in the \textsc{imfit} code) to estimate an initial set of parameters for the model. Subsequently, we ran the Levenberg-Marquardt algorithm in order to increase the best fitting model. 

During the fitting, we found the break radius of the thick disc to be larger than the semi-major axis of the galaxy. Therefore we changed the model for the thick disc to be a simple double-exponential disc without a break. This reduces the number of free parameters even further and, hence, the degeneracy of the fitted model. This is in agreement with \cite{2012ApJ...759...98C}, where it is shown that most thin discs are truncated (77 per cent for their sample of 70 edge-on galaxies, studied within the frame of the S$^4$G project), whereas thick discs are truncated in 31 per cent of the cases.

The results of the fitting are listed in Table~\ref{tab:IRAC_dec_res.tab}. As one can see, the thin disc is the dominant component with the truncation break radius at $R_{\text{b}}\approx2.7\,h_{R,\text{inn}}^{\text{t}}$, consistent with the results from \cite{2012ApJ...759...98C} (see their fig.~12). The thick disc is very extended and has very small flattening $h_z^{\text{T}} / h_R^{\text{T}} = 0.06$, with a radial scale length $\sim3.1$ times that of the thin disc.

The bulge contributes 19 per cent of the total flux, which confirms that IC\,2531 is a late type galaxy. However, the effective radius is quite large, which can be explained by the extended X-structure in the central part of the galaxy. The moderate \ser\ index $n\approx2.3$ suggests that the bulge lies in the overlapping area between pseudo-bulges and classical bulges \citep{2010ApJ...716..942F}. However, low bulge flattening suggests that the central component may appear to be a classical bulge rather than a pseudo-bulge. To disentangle this, a kinematic diagnostic of the inner region is required \citep{2012ApJ...754...67F}.

Using two sets of relative mass-to-light ratios $\Upsilon_\mathrm{T}/\Upsilon_\mathrm{t}=1.4$ and $\Upsilon_\mathrm{T}/\Upsilon_\mathrm{t}=2.4$ assumed in \cite{2011ApJ...741...28C}, we found the mass ratio
$M_\star^{\mathrm{T}}/M_\star^{\mathrm{t}}= $ 0.28 and 0.55 respectively. These values lie on the lower limit of the ranges for $M_\star^{\mathrm{T}}/M_\star^{\mathrm{t}}$ obtained in their study of thick and thin discs (see fig.~13 in \citealp{2011ApJ...741...28C}). This may indicate that the mass of the thick disc was slightly underestimated. However, this issue will not strongly affect our conclusions since the dust model parameters should not be substantially sensitive to the thick disc fraction to the total stellar model (see Sect.~\ref{sec:ini_model}).

\begin{figure}
\centering
\includegraphics[width=9.0cm, angle=0, clip=]{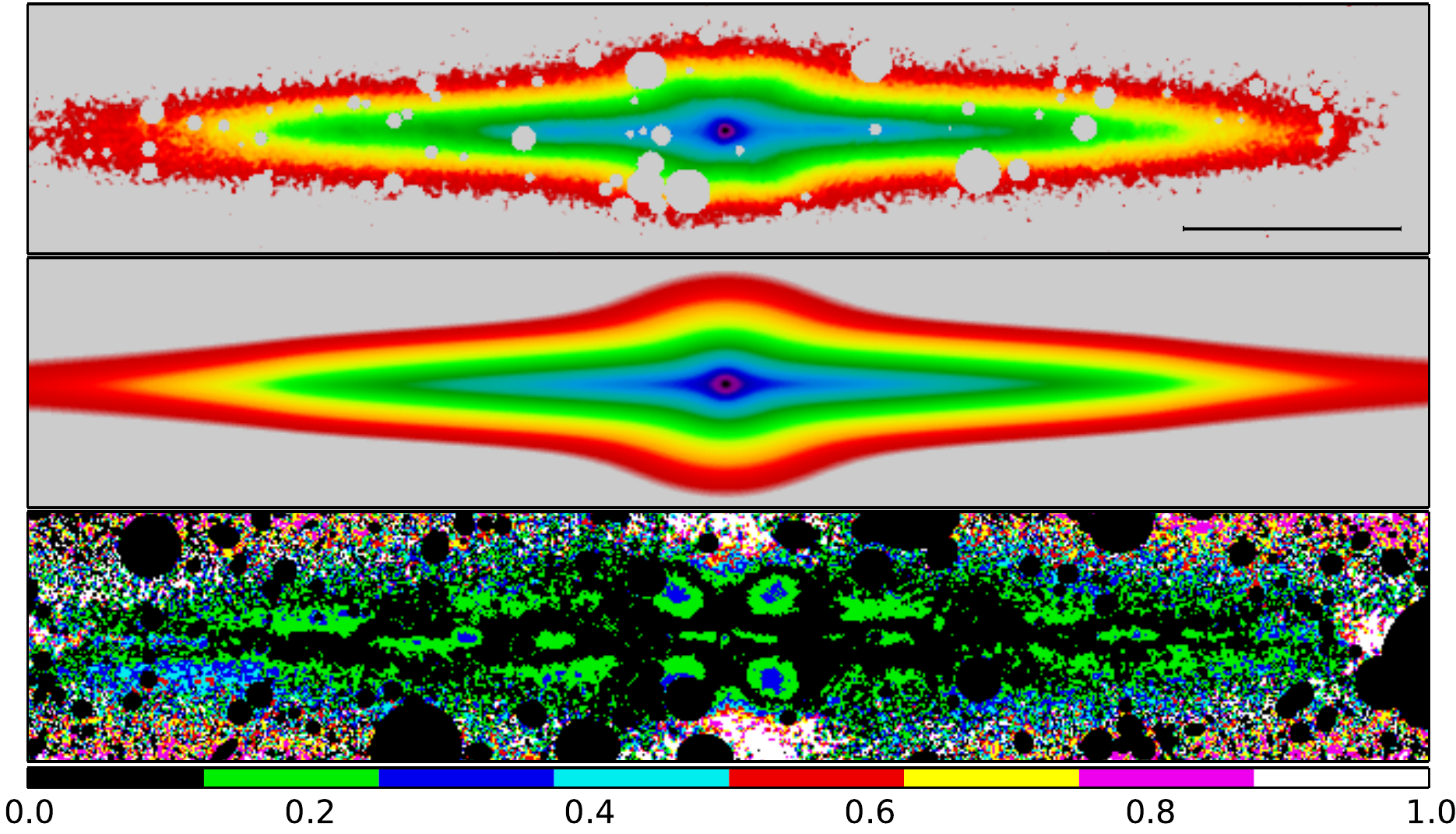}
\caption{The IRAC 3.6 $\mu$m image (top), the best fitting image (middle) and the residual image which indicates the relative deviation between the fit and the image in absolute values (bottom). The length of the bar on the top plot corresponds to 1 arcmin. The bottom color bar shows the scaling of the residual maps.}
\label{IC2531_irac_2d}
\end{figure}

\begin{figure*}[t]
\centering
\includegraphics[width=8.0cm, angle=0, clip=]{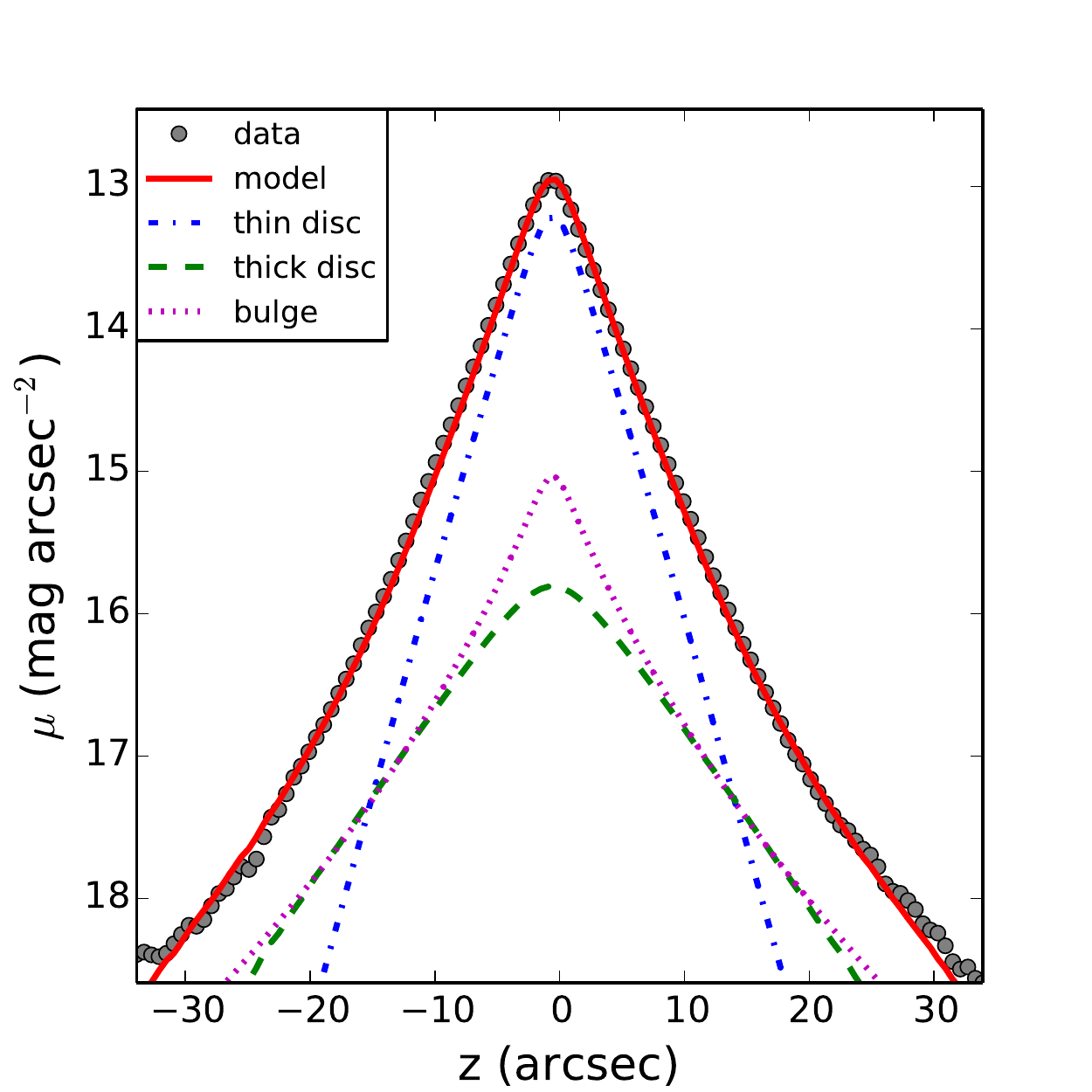}
\includegraphics[width=8.0cm, angle=0, clip=]{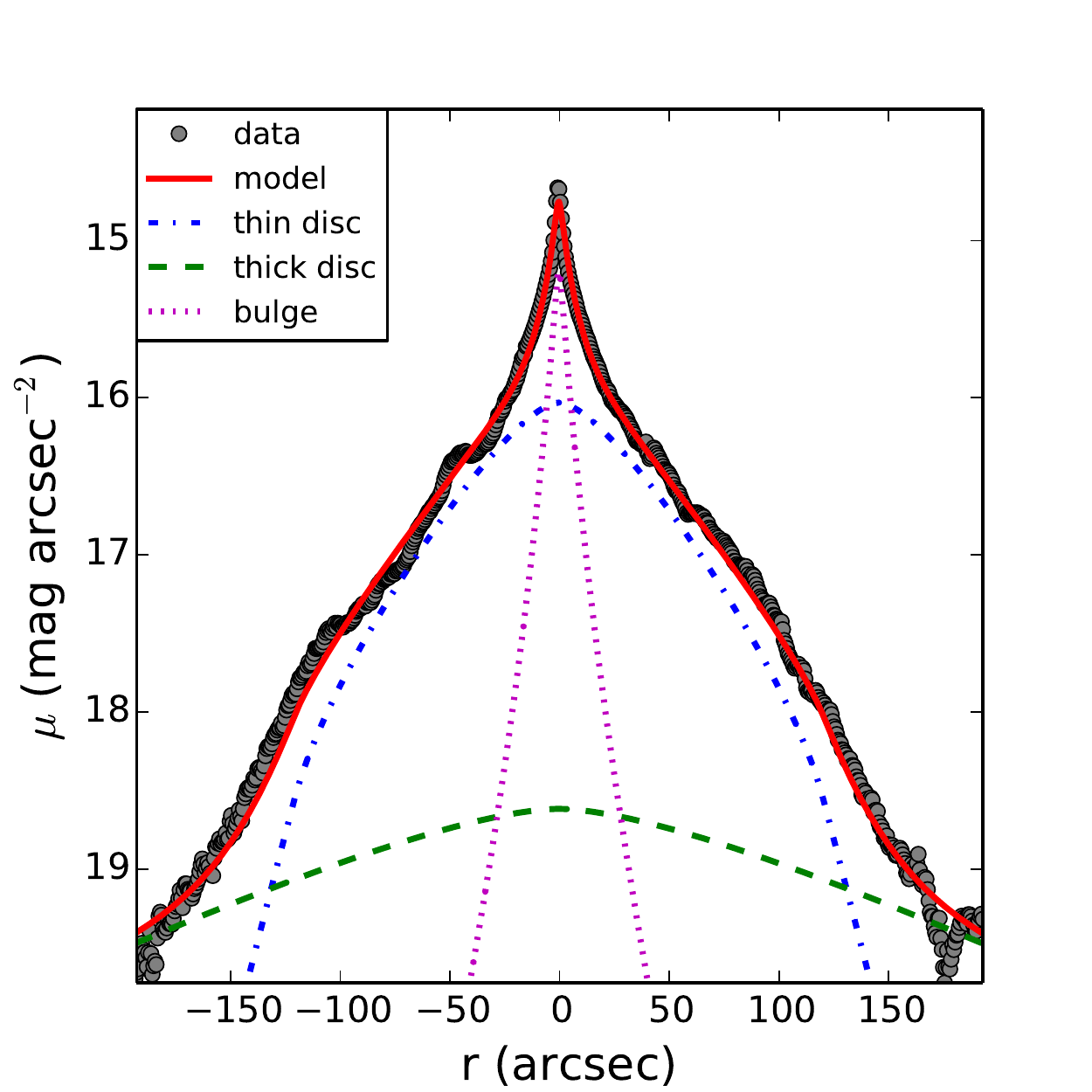}
\caption{Vertical profile summed up in the range of $|r|\geq50\arcsec$ where the fraction of the bulge is neglected (left) and horizontal profile summed up over all pixels in $z$-direction (right). {\it Comp\_1}, {\it Comp\_2} and {\it Comp\_3} in the legends represent the thin disc, the thick disc and the bulge respectively, as they are listed in Table~\ref{tab:IRAC_dec_res.tab}.}
\label{IC2531_irac_profiles}
\end{figure*}

In Fig.~\ref{IC2531_irac_2d} one can see the reference image, the model and the residual image for IC\,2531. Our model fit is very satisfactory, with the relative difference between the observation and the model below 25 per cent. Only the sharp X-structure in the centre slightly deviates from the model. The vertical and horizontal profiles in Fig.~\ref{IC2531_irac_profiles} are almost perfectly described by the model. We can see that the thick disc begins to dominate at large radii of the galaxy and has a very low surface brightness as compared to the thin disc.

In summary, we conclude that our galaxy model follows the observations very well and can be used for the subsequent analysis of the dust in the galaxy.

\begin{table}
\caption{Results of the decomposition of the IRAC 3.6\,$\mu\mathrm{m}$ image for IC\,2531.}
\label{tab:IRAC_dec_res.tab}
\centering
\resizebox{\columnwidth}{!}{
    \begin{tabular}{cccc}
    \hline
    \hline\\[-1ex]    
    Component & Parameter &  Value & Units \\[0.5ex]
    \hline\\[-0.5ex]
    1. Thin disc:&$h_\mathrm{R,inn}^\mathrm{t}$ & $8.0\pm0.54$ & kpc \\[+0.5ex]
    ({\it BrokenExponentialDisk3D})&$h_\mathrm{R,out}^\mathrm{t}$ & $3.33\pm0.58$& kpc \\[+0.5ex]
              &$h_\mathrm{z}^\mathrm{t}$ & $0.61\pm0.04$& kpc \\[+0.5ex]
              &$R_\mathrm{b}$ & $21.41\pm3.57$ & kpc  \\[+0.5ex] 
              &    $i$    & $89.6$ & deg \\[+0.5ex]
              &  $L_\mathrm{t}/L_\mathrm{tot}$ & $0.66\pm0.07$ & --- \\[+0.5ex]
    2. Thick disc:&$h_\mathrm{R}^\mathrm{T}$ & $24.87\pm0.77$& kpc \\[+0.5ex]
    ({\it ExponentialDisk3D})&$h_{z}^{T}$ & $1.57\pm0.18$ & kpc \\[+0.5ex]
              &    $i$    & $89.6$ & deg \\[+0.5ex]
              &  $L_\mathrm{T}/L_\mathrm{tot}$ & $0.15\pm0.03$ & --- \\[+0.5ex]
    3. Bulge:&$r_\mathrm{e,b}$ & $1.86\pm0.11$ & kpc \\[+0.5ex]
    (\ser) &$n_\mathrm{b}$ & $2.26\pm0.4$ & ---\\[+0.5ex]
    &$q_\mathrm{b}$ & $0.85\pm0.03$ & ---\\[+0.5ex]
    &  $L_\mathrm{b}/L_\mathrm{tot}$ & $0.19\pm0.05$ & --- \\[+0.5ex]
    \hline\\[-1ex]
    Total & $L_\mathrm{tot}$ & $-21.92\pm0.19$ & AB-mag \\[+0.5ex]
          & $M_\star$ & $7.91\pm1.69$ & $10^{10}\,\mathrm{M}_{\odot}$\\[+0.5ex]
    
    \hline\\[-0.5ex]
    \end{tabular}
}    
\parbox[t]{90mm}{ {\bf Notes.} The name of the \textsc{imfit} function is given in the parentheses under the name of the component. To find the total stellar mass, we relied on the mass-to-light ratio at 3.6\,$\mu\mathrm{m}$ of \cite{2012AJ....143..139E}.}
\end{table}

\subsection{{\sc{fitskirt}} fitting} 
\label{sec:fitskirt}

In order to build a realistic model for a dusty edge-on galaxy, we use the model obtained in Sect.~{\ref{sec:IRACdecomp}} to describe the stellar components of the galaxy, plus a dust disc represented by the double-exponential disc \citep[see e.g.][]{1999A&A.344.868X, 2010A&A...518L..39B, 2013A&A...556A..54V, 2013A&A...550A..74D, 2014MNRAS.441..869D}:
\begin{equation}
  \rho(R,z)
  =
  \frac{M_\mathrm{d}}{4\pi\,h_{R,\text{d}}^2\,h_{z,\text{d}}}
  \exp\left(-\frac{R}{h_{R,\text{d}}}\right)
  \exp\left(-\frac{|z|}{h_{z,\text{d}}}\right)\,,
\end{equation}
where $M_{\text{d}}$ is the total dust mass, and $h_{R,\text{d}}$ and $h_{z,\text{d}}$ are the radial scale length and vertical scale height of the dust respectively. The central face-on optical depth, often used as an alternative to express the dust content, is found as
\begin{equation}
  \tau_{\lambda}^\mathrm{f}
  \equiv
  \int_{-\infty}^{\infty}
  \kappa_{\lambda}\,\rho(0,z)\,{\mathrm{d}}z
  =
  \frac{\kappa_{\lambda}\,M_{\mathrm{d}}}{2\pi\,h_{R,\text{d}}^2}
\label{tauf}
\end{equation}
where $\kappa_{\lambda}$ is the extinction coefficient of the dust. The central edge-on optical depth can be found as $\tau_{\lambda}^\mathrm{e}=h_{R,\text{d}}/h_{z,\text{d}} \cdot \tau_{\lambda}^\mathrm{f}$. 

As discussed in the previous subsection, we fix the geometry of the different stellar components in the radiative transfer modelling, and we only fit the parameters of the dust disc, plus the centre alignment, the inclination angle $i$ and position angle $PA$ (even though we have aligned the galaxy plane with the horizontal axis, this alignment can still have a little error). Even though there is evidence that disc scale lengths are wavelength dependent \citep{1994A&AS..108..621P, 1996ASSL..209..523C, 1999A&A.344.868X}, the 3.6~$\mu$m surface brightness distribution best traces the stellar distribution of the galaxy, and we expect that our fixed geometrical model should be satisfactorily applicable at optical wavelengths as well. Note that we use the same geometrical parameters at all wavelengths for the dust disc and that only the luminosities of the bulge and the stellar discs are determined individually at each wavelength. This simulates the wavelength-dependent behaviour of their luminosity ratios \citep{2013A&A...550A..74D, 2014MNRAS.441..869D}. Note also that, at this stage, we do not make any assumptions about the characteristics of the stellar populations. We just fit the emission in every band individually and scale the emission without making assumptions about how the emission in different wavebands is linked. This approach allows as to significantly reduce the number of fitted parameters and, at the same time, use a complex model to describe the stellar and dust components of the galaxy. 

In order to describe our dust model, we should specify the optical properties of dust such as the absorption efficiency, the scattering efficiency, and the scattering phase function. We initially use the standard BARE-GR-S model described in \cite{2004ApJS..152..211Z} and implemented as in \citet{2015A&A...580A..87C}. This model consists of a mixture of polycyclic aromatic hydrocarbons (PAHs), graphite and silicate grains. The relative distributions of each grain component have been weighted to best match the extinction, abundances and emission associated with the Milky Way dust properties ($R_\mathrm{v}=3.1$). Note that the \textsc{skirt} code considers individual silicate, graphite and polycyclic aromatic hydrocarbon grain populations for the calculation of the dust emission spectrum, and that it fully takes into account the transient heating of PAHs and very small grains \citep{2015A&A...580A..87C}, using techniques described in \cite{1989ApJ...345..230G} and \cite{2001ApJ...551..807D}.

In order to fit the galaxy images described in Sect.~\ref{sec:data} with the model described above, we applied the oligochromatic radiative transfer fitting code {\sc{fitskirt}} \citep{2013A&A...550A..74D, 2014MNRAS.441..869D}. {\sc{fitskirt}} is a fitting code built around the 3D Monte Carlo radiative transfer code \textsc{skirt} \citep{2003MNRAS.343.1081B, 2011ApJS..196...22B, 2015A&C.....9...20C}\footnote{{\sc{SKIRT}} and {\sc{FITSKIRT}} are publicly available. The latest version can be downloaded from the official \textsc{skirt} web site \url{http://www.skirt.ugent.be}}. Our 3D model should reproduce the observed optical/NIR images of the galaxy, fully taking into account the effects of absorption and multiple anisotropic scattering by dust. \textsc{fitskirt} determines the best-fit parameters for the 3D distribution of stars and dust using genetic algorithm-based optimisation. Since the Monte Carlo method inevitably produces some Poisson noise, traditional optimisation tools such as Levenberg-Marquardt algorithm are not applicable in this case. Genetic algorithms, which have been successfully applied to a large range of global optimisation problems, can easily handle noisy objective functions \citep{2000ApJ...545..974M, 2003AJ....125.1958L, 2013A&A...550A..74D}. 

For this work, some new features have been added to the {\sc{fitskirt}} code. The code now can work with an arbitrary combination of stellar and dust components. A further improvement is that arbitrary PSF kernels can be provided, rather than the Gaussian kernels that were assumed in previous versions. The PSF is to be provided as a separate image, which can can be a cut out of an unsaturated star with high signal-to-noise ratio, or a model Gaussian or Moffat function. In case of images from space telescopes with very sophisticated PSF this new accurate PSF convolution is vital for retrieving reliable parameters of stellar components.

First, we performed oligochromatic fitting of the three images in $B$-, $V$- and $R$-band. After that, we repeated the same fitting but for seven bands: the same optical $B$-, $V$- and $R$-band images plus four near-IR images in the $J$, $H$, $K_\mathrm{s}$ and IRAC~3.6 passbands. The computations are done using the high performance cluster of the Flemish Supercomputer Center (Vlaams Supercomputer Centrum). In order to estimate the uncertainties of the parameters, we repeated the fitting five times for the models based on three as well as seven band images.  

The results of the fitting for both cases (best fits) are listed in Table~\ref{tab:fitskirt_res.tab}. As one can see, the model built upon the seven bands is significantly better constrained than the model based on only the optical images. Although the dust mass is almost the same, its uncertainty is very large in the first case, whereas the relative errors on each of the derived parameters for the second model are very small. In the first fit, the dust disc appears less extended and slightly thinner, which leads to a larger optical thickness. We will use the seven band model for our next analysis in Sect.~\ref{sec:simulations}.   

\begin{table}
\caption{Results of \textsc{fitskirt} fitting for IC\,2531.}
\label{tab:fitskirt_res.tab}
\centering
    \begin{tabular}{cccc}
    \hline
    \hline\\[-1ex]     
    Parameter &  $BVR$ & $BVRJHK3.6$ & Units \\
    \hline\\[-1ex] 
    $i$    & $89.15\pm0.51$ & $89.5 \pm 0.1$ & deg \\
    $h_\mathrm{R,d}$ & $6.17\pm 1.75$ & $8.44 \pm 0.29$ & kpc \\
    $h_\mathrm{z,d}$ & $0.20\pm 0.07$ & $0.25 \pm 0.01$ & kpc \\
    $M_\mathrm{d}$   & $4.13\pm 3.12$ & $4.08 \pm 0.22$ & $10^7~M_{\odot}$ \\
    $\tau_\mathrm{V}^{f}$ & $1.09\pm0.20$ & $0.57\pm0.01$ & --- \\
    $\tau_\mathrm{V}^{e}$ & $33.53\pm4.08$ & $19.26\pm0.18$ & --- \\[+0.5ex]
    \hline
    \end{tabular}
\end{table}

\begin{figure*}
\centering
\includegraphics[height=0.55\textwidth]{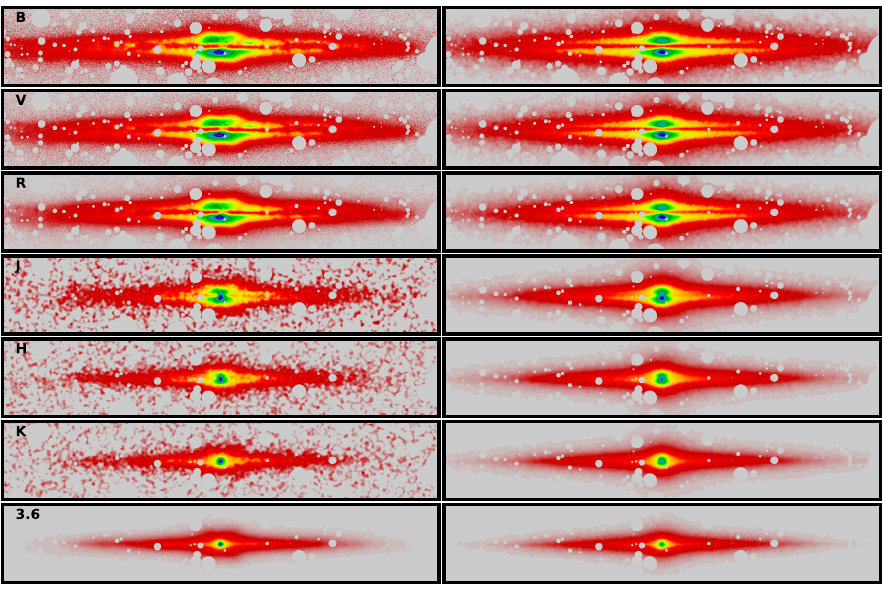}
\includegraphics[height=0.55\textwidth]{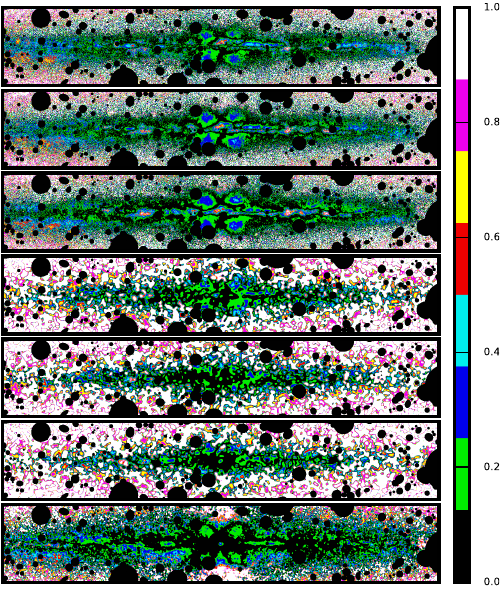}
\caption{Results of the oligochromatic \textsc{fitskirt} radiative transfer fits for IC\,2531. In each panel, the left-hand column represents the observed image, the middle column contains the corresponding fits in the same bands, the right-hand panel shows the residual images which indicate the relative deviation between the fit and the image.}
\label{IC2531_fitskirt_models}
\end{figure*}

The reference images (top left), models (top right) and the residual images (bottom) for the best fitting model are presented in Figure~\ref{IC2531_fitskirt_models}.  The \textsc{fitskirt} model reproduces the images quite accurately. One can see that the residuals show very little deviation in all passbands with slightly visible traces of the dust lane, especially where the disc has a perceptible warp and/or a possible spiral structure. The most prominent feature in the residual maps is the X-shaped pattern in the centre which cannot be fitted with our simple \ser\ bulge. 

The dust scale height of 0.25~kpc and the dust radial length of 8.44~kpc are in remarkable agreement with the average values $0.25$~kpc and $8.54$, respectively, found with \textsc{fitskirt} for 12 galaxies from the CALIFA sample \citep{2014MNRAS.441..869D}. However, the dust disc scales listed in Table~\ref{Disc_pars_authors} taken from \citet{1999A&A.344.868X} are 50 per cent larger than presented here. Also, the face-on optical depth in the $V$-band from our fit is a factor of two larger than found in the same work. However, the large edge-on optical depth in the $V$-band from our model is consistent with the average value of 19.09 for the sample from \cite{2014MNRAS.441..869D}. We should note that \citet{1999A&A.344.868X} used a different model with a single stellar disc and a de Vaucouleurs profile for the bulge, therefore a direct comparison in this case is impossible. In addition, while the \textsc{skirt} code uses a complete  treatment of absorption and multiple anisotropic scattering, \citet{1999A&A.344.868X} use an approximate radiative transfer algorithm in which higher-order scatterings are approximated \citep[see][for details]{1987ApJ...317..637K,1997A&A...325..135X}.

\section{Panchromatic radiative transfer modelling} 
\label{sec:simulations}

In this section, we extend the oligochromatic models obtained in Sect.~\ref{sec:fitskirt} to panchromatic models that can reproduce not only the images at optical and near-infared wavelengths, but also explain the entire UV--submm SED (Table~\ref{tab:Fluxes.tab}). In addition to that, we will be able to compare the predicted model images with corresponding observations at different wavelengths. This will help us to see how well our model can reproduce the visible structure of the galaxy and draw some conclusions from that.  

\subsection{Initial model} \label{sec:ini_model}

First, we directly use the earlier obtained oligochromatic \textsc{fitskirt} model (see Sect.~\ref{sec:fitskirt}) in order to predict the view of the galaxy in the entire UV to submm domain. This implies that both the properties for the stars and dust need to be set over this entire wavelength domain. For the stellar components, we assume a \cite{2003MNRAS.344.1000B} single stellar population SED with a \cite{2003PASP..115..763C} initial mass function and a solar metallicity ($Z=0.02$). We coarsely determined the mass-weighted ages of the stellar components using their de-reddened fluxes retrieved from the \textsc{fitskirt} modelling. For this, we used an adapted version of the SED fitting tool from \cite{2008MNRAS.386.1252H, 2009MNRAS.399.1206H}. We found that the ages of both the thin and the thick discs are about 5\,Gyr and about 8\,Gyr for the bulge.
For the Milky Way, the thick disc is primarily made of old and metal-poor stars compared to the thin stellar population \citep{1993ApJ...409..635R, 2000AJ....119.2843C}. This should be the case for IC\,2531 as well. However, since the thick disc is more extended in the vertical direction and faint in the dust disc plane (the thick disc-to-dust disc scale height ratio is larger than 6), its influence on the dust emission in the infrared should be minimal.

\begin{figure*}
\centering
\includegraphics[width=14.5cm, angle=0, clip=]{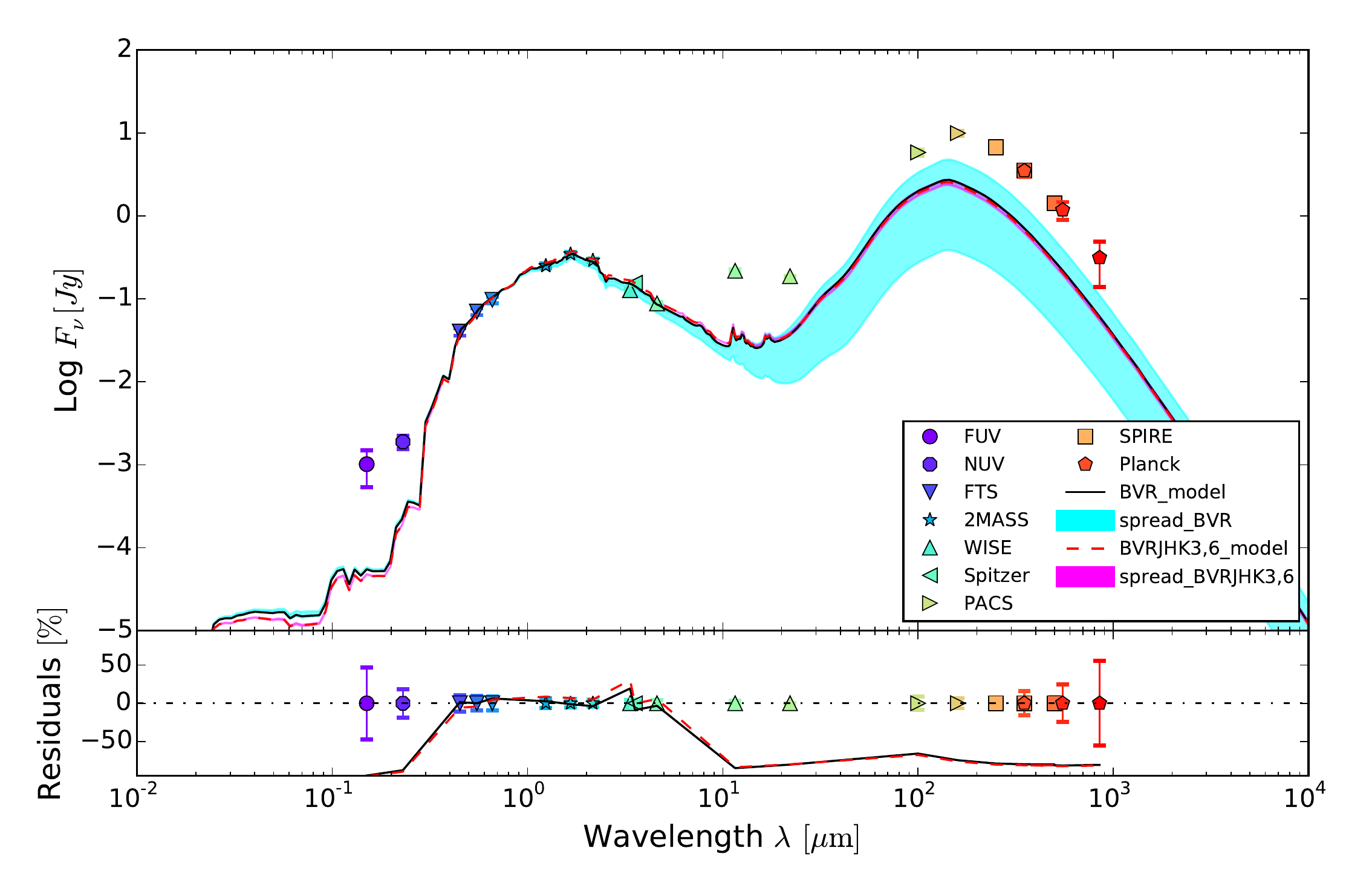}
\caption{The SEDs of IC\,2531 with the BARE-GR-S dust mixture based on the {\it{BVR}}-fit (black solid line) and the {\it{BVRJHK}}3.6-fit (red dashed line). The coloured markers with error bars correspond to the flux densities listed in Table~\ref{tab:Fluxes.tab}. The bottom panel below the SEDs show the relative residuals between the observed SEDs and the models. The cyan spread corresponds to the {\it{BVR}}-fit and the magenta spread relates to the {\it{BVRJHK}}3.6-fit. The spreads are plotted for the models within one error bar of the best oligochromatic fitting model parameters. }
\label{sed_trust_no_young_pop}
\end{figure*}

In Sect.~\ref{sec:fitskirt} we considered two models: one based only on the optical $B$, $V$ and $R$ images, and another one based on the optical and near-infrared images. The black line in Fig.~\ref{sed_trust_no_young_pop} is the predicted SED corresponding to the former model (the SED was normalised to match the $V$-band luminosity). The model reproduces the observed SED very well in the optical and near-infrared bands the model fluxes match the observations very well, even though it is only based on optical images. The agreement is much poorer outside this range: the model SED underestimates the observed flux densities in the UV and the dust emission in the MIR--submm region. The cyan coloured band indicates the spread of the model SED, generated according to the error bars of the dust parameters presented in Table~\ref{tab:fitskirt_res.tab}. Since some of the parameters are constrained relatively weakly, the spread is significant.

The red line in Fig.~\ref{sed_trust_no_young_pop} corresponds to the \textsc{fitskirt} model based on the {\it{BVRJHK}}3.6 photometry, with the well-constrained dust parameters. We will only refer to this oligochromatic model from now on since it is much better constrained and built on the near-infrared as well as the optical photometry. The panchromatic simulations are done in accordance with what was done above, except that the entire SED is normalised to match the IRAC~3.6~$\mu$m-band luminosity. The SEDs for both models have very similar behaviour and are almost indistinguishable from each other. However, the spread of possible flux density values for the {\it{BVRJHK}}3.6-based model (magenta colour) is barely visible. Nevertheless, the same underestimation of the UV and dust emission fluxes is present. 

\subsection{Models with an additional young stellar population} 
\label{sec:young_pop}

From Fig.~{\ref{sed_trust_no_young_pop}}, it is evident that we need an additional source of UV luminosity, i.e., a young stellar component, in order to match the observed SED of IC\,2531. As UV radiation is easily absorbed by interstellar dust, the addition of an additional young population will also affect the dust emission at MIR--submm wavelengths as well.


According to recent studies of the Milky Way \citep{2008ApJ...673..864J, 2011MNRAS.414.2446M}, the thickness of the thin disc is 200--300~pc, whereas the scale height of the young stellar disc which consists of OB-associations is about 50~pc \citep{1980ApJS...44...73B, 2016AstL...42....1B}. Moreover, the thickness of young stellar populations in spiral galaxies is similar or smaller than the dust disc scale height \citep{2013ApJ...773...45S, 2014ApJ...795..136S}. In our models, for the young stellar disc we assume a scale height of 100~pc and the same scale length of 8~kpc as for the thin disc. As shown by \citet{2014A&A...571A..69D} and \citet{2015MNRAS.451.1728D}, ranging the thickness of the young stellar disc between one third of the dust scale height to one dust scale height barely affects the resultant SED. Nevertheless, we will consider a possible influence of varying the thickness of the young stellar population on the model SED in Sect.~\ref{sec:themis}.  

The observed {\it GALEX} FUV image traces unobscured star formation regions of the galaxy. The stellar emission spectrum of these stars is described through a \textsc{starburst}99 SED template which represents a stellar population with a constant, continuous star formation rate (SFR) and an evolution up to 100 Myr \citep{1999ApJS..123....3L}. In this case, the initial mass function is a \cite{1955ApJ...121..161S} IMF with masses between 1 and 100\,M$_{\odot}$ and with solar metallicity. The luminosity of this component is constrained by the {\it GALEX} FUV flux density.

\begin{figure*}
\centering
\includegraphics[width=14.5cm, angle=0, clip=]{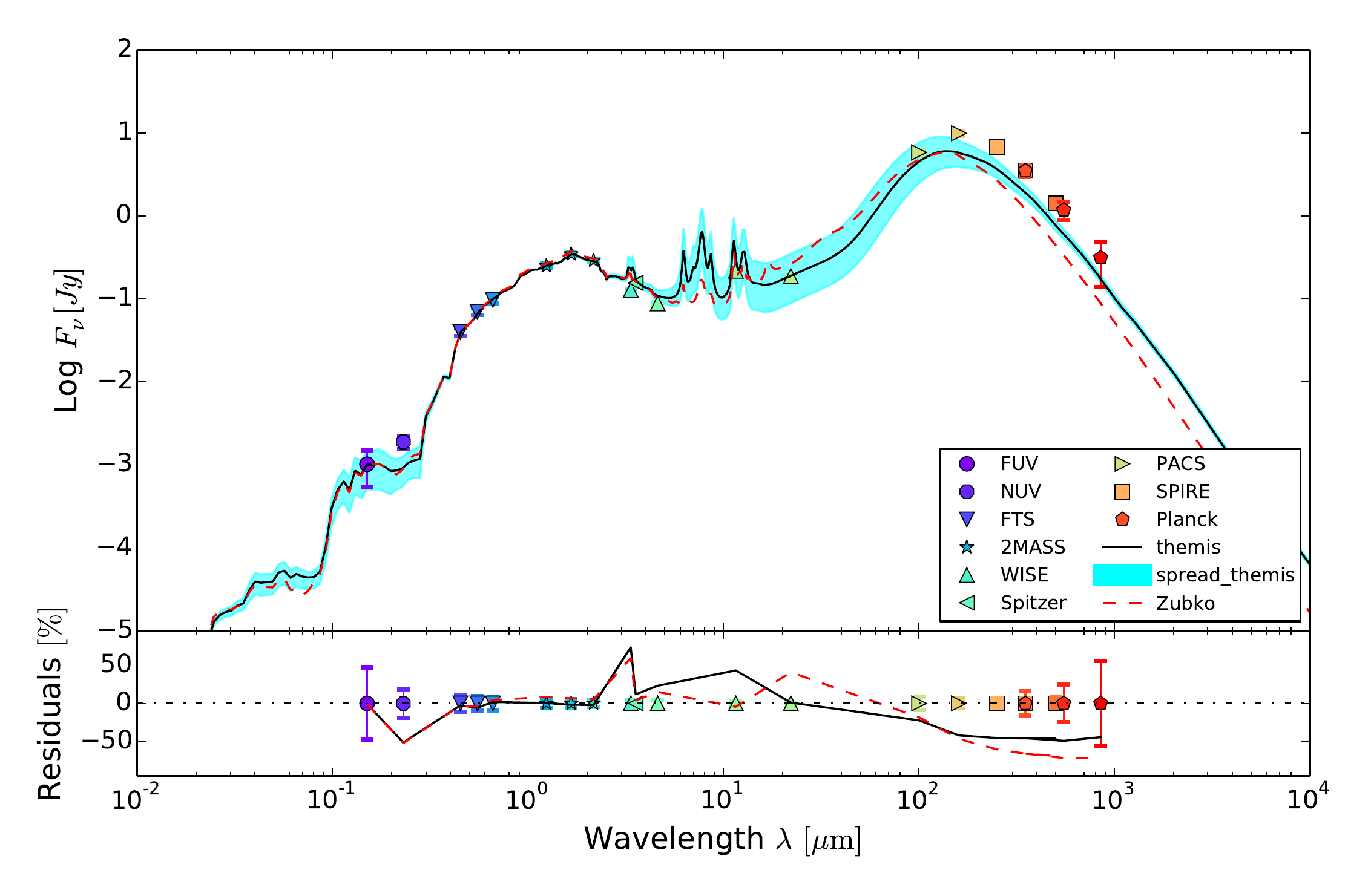}
\caption{The SED of IC\,2531 based on the BARE-GR-S (the dashed line) and THEMIS (the solid line) dust mixture, with an additional young stellar component. The coloured symbols with error bars correspond to the flux densities listed in Table~\ref{tab:Fluxes.tab}. The cyan band represents the spread in the MIR and FIR emission of the THEMIS model with different values of the scale height of the young stellar population (see text). The bottom panel below the SED show the relative residuals between the observed SED and the models.}
\label{sed_young_pop}
\end{figure*}

The simulated SED of this updated model is shown in Fig.~\ref{sed_young_pop} (red dashed line). As one can see, the contribution of this young stellar population is negligible in the optical and NIR bands, and, on the other hand, describes the MIR/FIR region significantly better compared to the previous model without the inclusion of the young stellar population. The {\it WISE} $W3$ flux is perfectly reproduced, whereas the PACS 100~$\mu$m flux is about 25 per cent underestimated. At the same time, the model overestimates the observed WISE 22~$\mu$m flux density by more than 70 per cent, whereas it significantly underestimates the observed flux densities beyond 100 $\mu$m. We should also point out that the model appears to underestimate the observed NUV flux as well. This could be linked to our choice of the dust model: while extinction curves of different galaxies tend to agree well at optical wavelengths, they show quite some diversity in the UV, in particular around the 2175~\AA\ bump \citep[e.g.,][]{2003ApJ...594..279G}.

\subsection{THEMIS dust model} \label{sec:themis}

In Sect.~\ref{sec:young_pop} we showed that the \citet{2004ApJS..152..211Z} BARE-GR-S model overestimates the MIR and underestimates the submm flux densities of IC\,2531. We decided to verify whether these discrepancies are related to the adopted dust model. To this end, we replace the BARE-GR-S dust model with the so-called THEMIS dust model presented by \citet{2013A&A...558A..62J} and \citet{2014A&A...565L...9K, 2015A&A...579A..15K}. This model is built completely on the basis of interstellar dust analogue material synthesised, characterised and analysed in the laboratory. In this model, there are two families of dust particles: amorphous carbon and amorphous silicates. For the silicates, it is assumed that 50 per cent of the mass is amorphous enstatite, and that the remaining half is amorphous forsterite.

The same simulations as in the previous Sect.~\ref{sec:young_pop} were carried out. We used exactly the same model for the stellar components and the dust disc parameters, with the additional young stellar population. The only difference was the dust composition, which was changed to the THEMIS dust model. In Fig.~\ref{sed_young_pop} (the black solid line), one can see that the model reproduces the MIR/FIR SED region better than that with the BARE-GR-S dust implementation, especially at {\it WISE} 22~$\mu$m, SPIRE and {\it Planck} wavelengths. However, the comparison between the model and the observations is still not fully satisfactory. A substantial FIR excess remains: the model dust emission  appears to be underestimated by a factor of 2.

The cyan spread in Fig.~\ref{sed_young_pop} is related to the models with maximal and minimal dust emission and takes into account two aspects. First, it includes the uncertainty of the measured FUV flux density. Second, we varied the disc thickness of the young stellar population from 50~pc to 150~pc (in fact, we found this variance of the disc thickness has a minor affect on the SED). It is seen that the difference in the SED is notable in the MIR/FIR region up to $\lambda\la200\,\mu$m, whereas for the cold dust the difference is negligibly small.

It is interesting to compare the SEDs for the THEMIS and BARE-GR-S dust models which are both presented in Fig.~\ref{sed_young_pop}. We can see the main difference in the MIR-submm domain. The THEMIS model overestimates the {\it WISE} $W3$ flux and fully restores the {\it WISE} $W4$ flux, whereas for the BARE-GR-S model we see the contrary result at the same wavelengths. The THEMIS model (together with its spread) accounts for the PACS $100~\mu$m flux density and produces more dust emission in the FIR/submm region compared to the BARE-GR-S model. We will discuss this result in Sect.~\ref{sec:discussion}.

\subsection{Comparing the observed and simulated images} \label{sec:maps}

\begin{figure*}
\centering
\includegraphics[width=\textwidth, angle=0, clip=]{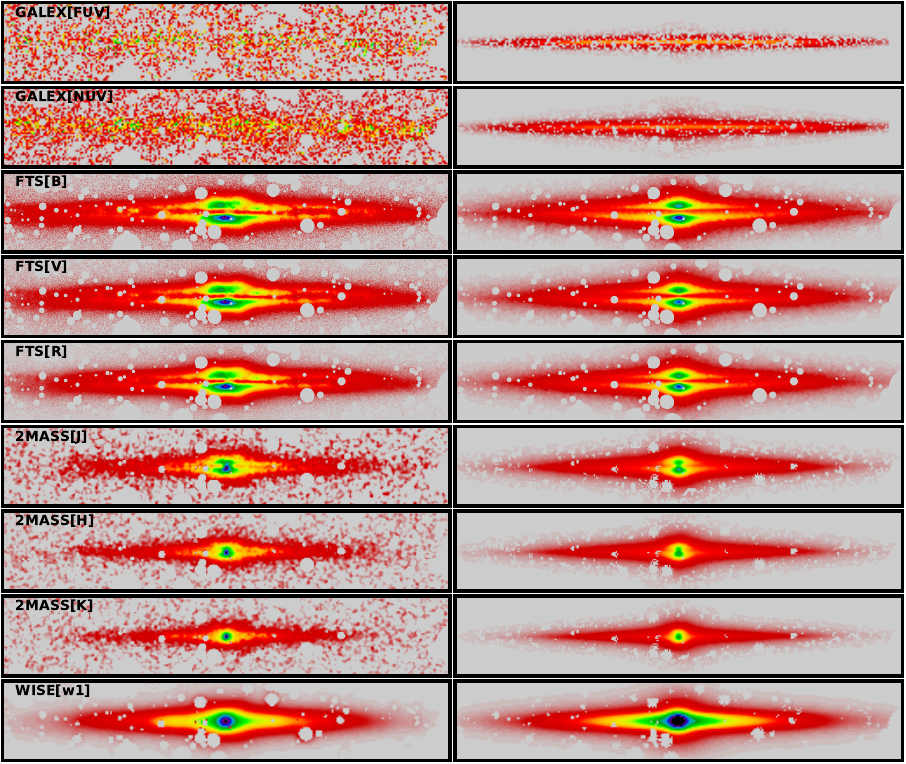}
\caption{Comparison between the observations (left) and model images (right).}
\label{map1}
\end{figure*}

\addtocounter{figure}{-1}
\begin{figure*}
\centering
\includegraphics[width=\textwidth, angle=0, clip=]{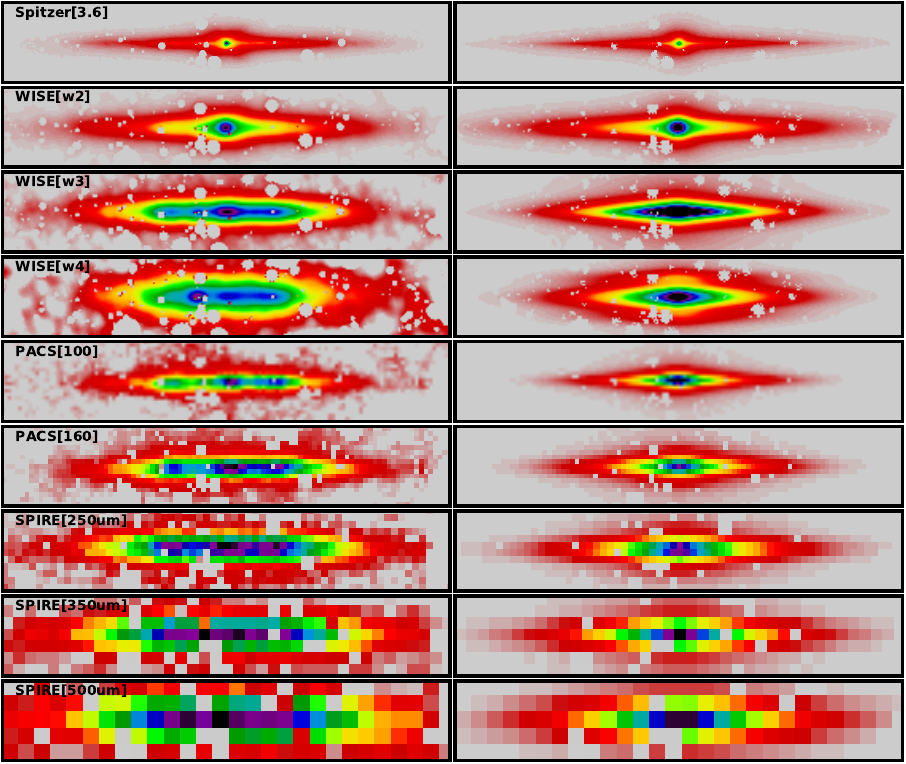}
\caption{(continued)}
\end{figure*}

In Fig.~\ref{map1}, we compare a set of images corresponding to the THEMIS dust model described in Sect.~\ref{sec:themis} to the reference images in different bands across the entire wavelength range, from the UV to the submm. The modelled images have been convolved with the appropriate point spread function: for FTS, 2MASS and IRAC 3.6~$\mu$m we used the PSFs built in Sect.~\ref{sec:data}, for the other bands we used the kernels provided by \cite{2011PASP..123.1218A}. 

The {\it GALEX} FUV image is too noisy to be compared, and we provide it here only for completeness. The {\it GALEX} NUV image is better comparable, but still the deepness of the image is not sufficient to do a reliable analysis of it. Nevertheless, we can see that the 2D profile of the simulated image resembles the observed one, though the vertical surface brightness distribution in the observed image seems wider. In an attempt to remedy the situation with the low signal-to-noise ratio of the observed image, we fitted the vertical profile summed over all radii with an exponential function and estimated its scale height $h_\mathrm{z}=531\pm7$~pc, whereas for the simulated NUV image we obtained $h_\mathrm{z}=215\pm3$~pc. This UV excess can be explained by the presence of another, more vertically extended young stellar component. Alternatively, it can be diffuse scattered light produced by a radially extended PSF \citep{2014A&A...567A..97S} or scattered emission from a vertically extended dust distribution \citep[see e.g][]{2014ApJ...789..131H, 2015ApJ...815..133S, 2016A&A...587A..86B}. 

The next set of simulated images, from the FTS {\it{B}} band to the {\it WISE} $W2$ band, are in very good agreement with the observations. However, the periphery of the galaxy disc exhibits prominent flaring (an apparent increase of the disc thickness with radius). This can be due to real physical flaring or the projection of disc warps, or the specific orientation of spiral arms towards the observer. 

\begin{figure}
\centering
\includegraphics[width=9cm, angle=0, clip=]{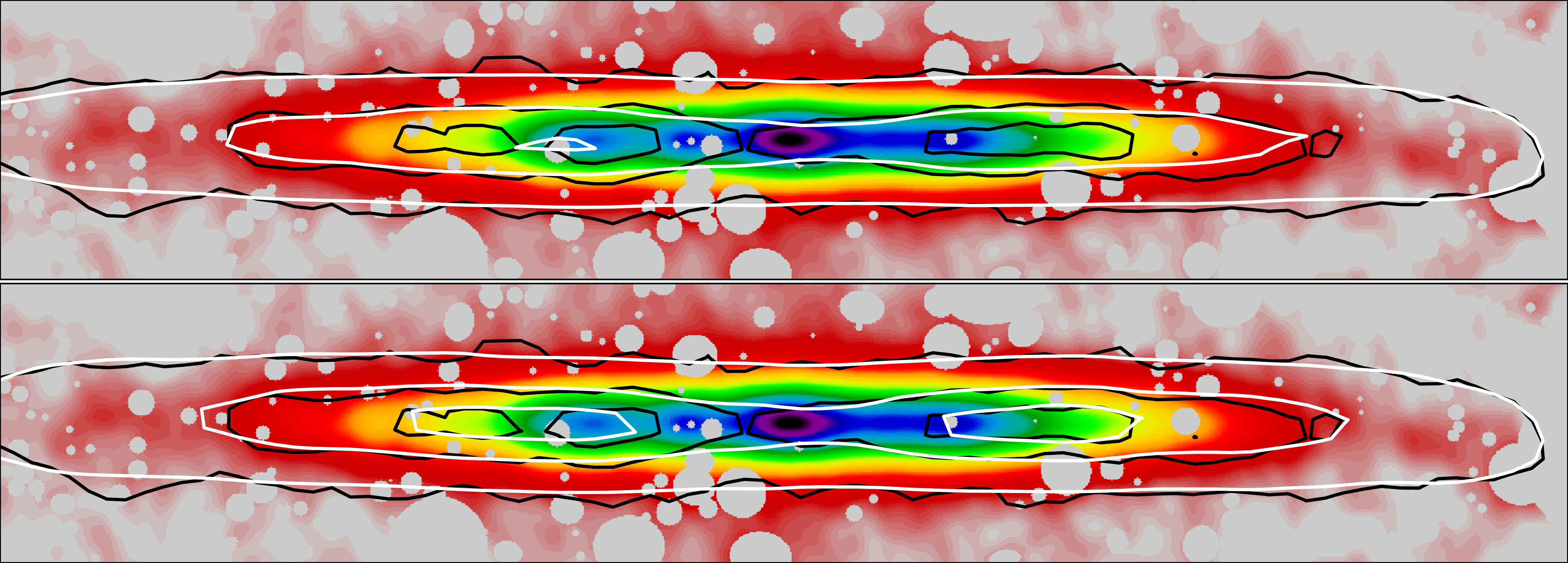}
\caption{Comparison of two models from \cite{2015A&A...582A..18A} without (top plot) and with arcs mimicking the effect of spiral arms (bottom plot), built on the basis of the kinematical information analyzed in their work. For both plots, the {\it WISE} $W3$ image is overplotted with the observed H{\sc{i}} contours (black) and the contours from a model (white). Contour levels are $1.49\times10^{21}$, $4.475\times10^{21}$ and $6.56\times10^{21}$ atoms~cm$^{-2}$.}
\label{spirals_HI}
\end{figure}

In the images from {\it WISE} $W3$ to PACS 100~$\mu$m, the signature of this spiral structure is visible.
In Fig.~\ref{spirals_HI} one can see the comparison of two models for IC\,2531 taken from \cite{2015A&A...582A..18A}. Both the top and bottom plots represent the {\it WISE} $W3$ image, overplotted with the observed H{\sc{i}} contours (black) and the white contours for a model that does not contain spiral arms (top) and a toy model that includes arcs  to mimic the effect of spiral arms in the position-velocity diagram (bottom). This comparison shows that spiral arms could indeed explain the overdensities seen in the {\it WISE} $W3$ image. Spiral arms are probable also present in the other PACS and SPIRE images (the dust emission component is more radially extended as compared to the modelled one), but the spatial resolution at those wavelengths is not sufficient to draw strong conclusions. Also, one can see that the profile of the observed structure at these wavelengths has a rounder shape in comparison with the model that looks more discy. Again, this discrepancy may be related to the presence of spiral structure, which can widen the vertical distribution of the surface brightness along the major axis of the galaxy. As has been mentioned before, the SPIRE images appear about twice more luminous than the simulations.

In general, our simulations seem to match the observations reasonably well, except for the FIR/submm wavelengths where the galaxy geometry is more complex and the total fluxes are underestimated by a factor of about 2.

\section{Discussion} 
\label{sec:discussion}

\subsection{The need for oligochromatic fitting} \label{sec:olifitting}

As was shown in Sect.~\ref{sec:ini_model}, the uncertainties on the dust disc parameters can be dramatically affected by the data which are taken into account: the optical+NIR based data model is much more constrained than the model where only the optical images are fitted. This is similar to what was found by \cite{2014MNRAS.441..869D}, where they validated the oligochromatic fitting procedure on simulated images. They showed that oligochromatic models have smaller error bars on most of the fitted model parameters than the individual monochromatic models. In our case of a real galaxy, we can see that the uncertainty of the model can be substantial even if we do oligochromatic fitting of optical data in several bands. For a more precise model, NIR data is needed. Evidently, the more data we use, on a larger scale of wavelengths, the better constrained our model will be, which, in principal, should predict a more consistent SED behaviour at other wavelengths.

The result that adding NIR images significantly constrains the dust model should be discussed. First, let us turn to the work by \cite{2014MNRAS.441..869D}, where they analysed a sample of 12 edge-on spiral galaxies. For this, they used $g$-, $r$-, $i$- and $z$-band images from the Sloan Digital Sky Survey \citep{2014ApJS..211...17A} to perform \textsc{fitskirt} oligochromatic fitting with a model consisting of a double exponential stellar disc, a bulge and an exponential disc for the dust component. In general, the dust parameters they retrieved from their fitting are quite well constrained. For a few galaxies, however, the error bar on the dust mass was up to 50 percent. Such large uncertainties can be due to various reasons. First, the simple stellar model, consisting of a bulge and an exponential disc, does not accurately describe the observed morphology of the galaxy. Second, using an exponential disc model can be a too simplified model if dust in the galaxy is distributed in a more complex way. Third, the large uncertainties of the dust model can be explained by an insufficient angular resolution of the images used for the galaxy fitting.

In the case of IC\,2531, it is unlikely that any of the above cause the large uncertainties on the dust parameters for the {\it{BVR}} model. We have seen that our stellar model describes the observations very well. Also, the exponential disc model for the dust seems to be reasonable. The emission coming from an extended structure in the PACS and SPIRE images is likely related to spiral structure. In addition, the {\it{BVR}} images used for the {\sc{fitskirt}} modelling are sufficiently large compared to the images of the galaxies fitted in \citet{2014MNRAS.441..869D}.

A possible explanation of the poorly constrained dust disc for the {\it{BVR}}-based model might be due to the fact that we have set the geometry of the stellar components for the optical images based on the observed IRAC 3.6~$\mu$m images. The actual geometry of the stellar components at optical wavelengths could be different, as stellar populations of different ages dominate at different wavelengths. As was mentioned in Sect.~\ref{sec:IRACdecomp}, however, direct fitting of galaxy images with a complex model consisting of three stellar components and a dust disc, is a too ambiguous computational task. Our approach of using an near-infrared galaxy image for building the stellar model (and adopting it at all wavelengths) seems to be the only possible solution to avoid this issue.

There is another concern related to the dust model constraints. Since NIR images are not strongly affected by dust extinction, it is not evident why adding NIR images in fitting should constrain the dust model parameters. Nevertheless, if we look at the SEDs for the {\it{BVR}} and {\it{BVRJHK}}3.6 models (see Fig.~\ref{sed_trust_no_young_pop}), we can see that the spread for the former is negligible in the {\it{J}}-band and becomes gradually larger at longer wavelengths: the uncertainty of the flux density at 3.6~$\mu$m is about 20 per cent of the observed value, whereas it is about 9 per cent for the {\it{J}}-band. Moreover, the significant spread for the {\it{BVR}}-based model at 3.6~$\mu$m shows that, in fact, the dust model may non-negligibly affect the total flux in this band. Indeed, even rather opaque models (up to $\tau^{\text{f}}_{3.6}=0.19$) are possible within the range of dust parameters. However, if we fit optical and NIR data, our model SED becomes significantly better constrained at the considered wavelengths, which subsequently constrains the dust model and the predicted SED at other wavelengths.    


\subsection{The dust model} 
\label{sec:dustmodel}

In Sect.~\ref{sec:themis}, we compared two physical dust models: the standard BARE-GR-S model from \citet{2004ApJS..152..211Z} versus the THEMIS dust model by \citet{2013A&A...558A..62J} and \citet{2014A&A...565L...9K, 2015A&A...579A..15K} . It is apparent that including the {\it Herschel} PACS and SPIRE data should set constraints on both the warm and cold dust components. However, these two different dust models result in different dust emission SEDs: compared to the BARE-GR-S model, the THEMIS model produces less mid-infrared emission, but more emission at FIR and submm wavelengths. This difference can be explained by the fact that the dust grains in these models have different optical and calorimetric properties, which apparently leads to, on average, cooler dust for the THEMIS model.

\cite{2015A&A...580A.136F} compared several dust models in reproducing dust emission and extinction for Galactic diffuse ISM observations from the {\it Planck}, IRAS and SDSS surveys. They found that the THEMIS model better matches observations than the \cite{2007ApJ...657..810D} and \cite{2011A&A...525A.103C} models. Also, it produces more emission in the far-infrared and submillimeter domain, while ``astronomical silicate'' and graphite are not emissive enough in that region. 

All in all, we can conclude that the excess of the SED (data -- model) strongly depends on the adopted dust mixture model. The question of which model is, from a physical point of view, the more appropriate to use is still debatable, though for IC\,2531 the THEMIS model seems to provide the results which are in better agreement with the observations. We will compare these dust mixtures for the remaining {\it{HER}}OES galaxies in our next study. Obviously, the choice of the dust model can be one of the sources of the dust energy balance problem.

\subsection{The dust energy balance and the HEROES sample} 
\label{sec:energybalance}

The dust energy balance problem for IC\,2531 is not completely solved in this paper. However, we have shown that the dust mixture model plays an important role in shaping the model SED. 

With time, this problem seems to become more complicated as several possible solutions are being proposed, but none of them can be considered as dominant. A possible explanation is the presence of additional dust distributed in the form of a second inner disc that hardly contributes to the galaxy attenuation, but still provides a considerable fraction of the FIR emission \citep{2000A&A...362..138P, 2001A&A...372..775M,2011A&A...527A.109P}. As recently shown by \citet{2015A&A...576A..31S}, complex asymmetries and the inhomogeneous structure of the dust medium may help to explain the discrepancy between the dust emission predicted by radiative transfer models and the observed emission in energy balance studies for edge-on spiral galaxies. Another source of the infrared excess can be completely obscured star forming regions \citep[see e.g.][]{2014A&A...571A..69D, 2015MNRAS.451.1728D} It is quite possible that this problem encompasses several mechanisms which may work at the same time.

As advocated by \citet{2015MNRAS.451.1728D} and \citet{2015A&A...576A..31S}, the best way to discriminate between the various possible explanations (or estimate their contribution) is to select a sample of galaxies with a rich data set and model them in a uniform way. Apart from a detailed study of IC\,2531, an interesting and well-studied galaxy by itself, a second goal of this paper was to develop a modelling approach that can be applied to a larger set of galaxies. We will apply the approach presented in this work to all {\it HER}OES galaxies. Stellar components for each object will be determined from near-infrared images, taking into account thin and thick discs if appropriate. The dust component will be modelled with \textsc{fitskirt}, by simultaneously analyzing images in a number of optical and near-infrared bands. This analysis will provide us with a 3D model for each galaxy, including their geometry and masses of stellar and dust components. Taking into account additional data ranging from UV to submm wavelengths, we will be able to construct model SEDs, compare them with observations and make a quantitative assessment of the dust energy balance problem for each galaxy. Also, we will investigate the detailed distribution of dust in all {\it HER}OES galaxies and obtain some statistics of dust properties on the basis of the performed analysis, which can be compared to previous works. Models of stellar components will be used in our next study of dynamical properties of the {\it HER}OES galaxies.

\section{Summary}
\label{sec:summary}

We have explored the distribution of stars and dust, and the dust energy balance of the edge-on spiral galaxy IC\,2531 using the available observations from UV to submm wavelengths. We have expanded the modelling method of \citet{2015MNRAS.451.1728D} to a three-step approach. We first construct a detailed parametric model for the stellar components, based on a decomposition of the IRAC 3.6~$\mu$m image. Then, we carry out an oligochromatic radiative transfer fitting of the optical plus near-infrared data, in order to retrieve the dust content parameters. Finally, we expand this model to a panchromatic model and predict the resulting SED over the entire UV--submm wavelength domain. 

The main findings of this article can be summarised as follows.
\begin{enumerate}
\item{We manage to build a radiative transfer model that accurately reproduces the observed images of IC\,2531 at optical and near-infrared wavelengths. The structural parameters are in good agreement with previous studies of other edge-on spiral galaxies.}
\item{Even when adding an additional young stellar disc to the model to fit the UV flux densities, the radiative transfer model underestimates the observed far-infrared emission by a factor of a few.}
\item {Adding near-infrared imaging to the optical data in the oligochromatic fitting significantly constrains the parameters of the dust distribution.}
\item {Different dust models as, for example, the BARE-GR-S and THEMIS dust models used in this work, produce slightly different SEDs. This might be important to determine the nature of the dust energy balance problem.}
\end{enumerate}

\begin{acknowledgements}
A.V.M. is a beneficiary of a postdoctoral grant from the Belgian Federal Science Policy Office, and also expresses gratitude for the grant of the Russian Foundation for Basic Researches number 14-02-00810 and 14-22-03006-ofi. 

F.A., M.B., I.D.L. and S.V. gratefully acknowledge the support of the Flemish Fund for Scientific Research (FWO-Vlaanderen). 

M.B. acknowledges financial support from the Belgian Science Policy Office (BELSPO) through the PRODEX project “Herschel-PACS Guaranteed Time and Open Time Programs: Science Exploitation” (C90370). 

T.M.H. acknowledges the CONICYT/ALMA funding Program in Astronomy/PCI Project No:31140020. 

J.V. acknowledges support from the European Research Council under the European Union’s Seventh Framework Programme (FP/2007-2013)/ ERC Grant Agreement nr. 291531. 

This work was supported by CHARM and DustPedia. CHARM (Contemporary physical challenges in Heliospheric and AstRophysical Models) is a phase VII Interuniversity Attraction Pole (IAP) program organized by BELSPO, the BELgian federal Science Policy Office. DustPedia is a collaborative focused research project supported by the European Union under the Seventh Framework Programme (2007-2013) call (proposal no. 606847). The participating institutions are: Cardiff University, UK; National Observatory of Athens, Greece; Ghent University, Belgium; Université Paris Sud, France; National Institute for Astrophysics, Italy and CEA (Paris), France.

The Faulkes Telescopes are maintained and operated by Las Cumbres Observatory Global Telescope Network. We also thank Peter Hill and the staff and students of College Le Monteil ASAM (France), The Thomas Aveling School (Rochester, England), Glebe School (Bromley, England) and St David’s Catholic College (Cardiff, Wales). 

This work is based in part on observations made with the Spitzer Space Telescope, which is operated by the Jet Propulsion Laboratory, California Institute of Technology under a contract with NASA. 

This publication makes use of data products from the Two Micron All Sky Survey, which is a joint project of the University of Massachusetts and the Infrared Processing and Analysis Center/California Institute of Technology, funded by the National Aeronautics and Space Administration and the National Science Foundation. 

This publication makes use of data products from the Wide-field Infrared Survey Explorer, which is a joint project of the University of California, Los Angeles, and the Jet Propulsion Laboratory/California Institute of Technology, funded by the National Aeronautics and Space Administration. 

This study makes use of observations made with the NASA Galaxy
Evolution Explorer. GALEX is operated for NASA by
the California Institute of Technology under NASA contract
NAS5-98034.

The Herschel spacecraft was designed, built, tested, and
launched under a contract to ESA managed by the Herschel/Planck
Project team by an industrial consortium under the overall responsibility
of the prime contractor Thales Alenia Space (Cannes), and
including Astrium (Friedrichshafen) responsible for the payload
module and for system testing at spacecraft level, Thales Alenia
Space (Turin) responsible for the service module, and Astrium
(Toulouse) responsible for the telescope, with in excess of a hundred
subcontractors.

SPIRE has been developed by a consortium of institutes led
by Cardiff Univ. (UK) and including: Univ. Lethbridge (Canada);
NAOC (China); CEA, LAM (France); IFSI, Univ. Padua (Italy);
IAC (Spain); Stockholm Observatory (Sweden); Imperial College
London, RAL, UCL-MSSL, UKATC, Univ. Sussex (UK); and
Caltech, JPL, NHSC, Univ. Colorado (USA). This development
has been supported by national funding agencies: CSA (Canada);
NAOC (China); CEA, CNES, CNRS (France); ASI (Italy); MCINN
(Spain); SNSB (Sweden); STFC, UKSA (UK); and NASA (USA).

We acknowledge the use of the ESA Planck Legacy Archive.

Some of the data presented
in this paper were obtained from the Mikulski Archive for
Space Telescopes (MAST). STScI is operated by the Association
of Universities for Research in Astronomy, Inc., under
NASA contract NAS5-26555. Support for MAST for nonHST
data is provided by the NASA Office of Space Science
via grant NNX13AC07G and by other grants and contracts.

This research makes use of the NASA/IPAC Extragalactic Database (NED) which is operated by the Jet Propulsion Laboratory, California Institute of Technology, under contract with the National Aeronautics and Space Administration, and the LEDA database (http://leda.univ-lyon1.fr).
 \\
\end{acknowledgements}

\bibliographystyle{aa} 
\bibliography{ic2531}

\begin{thebibliography}{}
\makeatletter
\def\@urlcharsother{\let\do\@makeother\do\\\do\$\do\&\do\#\do\^\do\_\do\%\do\~}
\def\doi{\begingroup\@urlcharsother\@ifnextchar[{\@doi}{\@doi[]}}
\def\@doi[#1]#2{\def\@tempa{#1}\ifx\@tempa\@empty\href{http://dx.doi.org/#2}{doi:#2}\else\href{http://dx.doi.org/#2}{#1}\fi\endgroup}
\def\eprint#1#2{\@eprint#1:#2::\@nil}
\def\@eprint@arXiv#1{\href{http://arxiv.org/abs/#1}{{\tt arXiv:#1}}}
\def\@eprint@dblp#1{\href{http://dblp.uni-trier.de/rec/bibtex/#1.xml}{dblp:#1}}
\def\@eprint#1:#2:#3:#4\@nil{\def\@tempa{#1}\def\@tempb{#2}\def\@tempc{#3}\ifx\@tempc\@empty\let\@tempc\@tempb\let\@tempb\@tempa\fi\ifx\@tempb\@empty\def\@tempb{arXiv}\fi\@ifundefined{@eprint@\@tempb}{\@tempb:\@tempc}{\expandafter\expandafter\csname
  @eprint@\@tempb\endcsname\expandafter{\@tempc}}}
\def\mniiiauthor#1#2#3{\@ifundefined{mniiiauth@#1}{\global\expandafter\let\csname
  mniiiauth@#1\endcsname\null #2}{#3}}
\makeatother

\bibitem[Ahn et al.(2014)]{2014ApJS..211...17A} Ahn, C.~P., Alexandroff, R., Allende Prieto, C., et al.\ 2014, \apjs, 211, 17 

\bibitem[Allaert et al.(2015)]{2015A&A...582A..18A} Allaert, F., Gentile, G., Baes, M., et al.\ 2015, \aap, 582, A18 

\bibitem[\protect\citeauthoryear{{Alton}, {Xilouris}, {Misiriotis}, {Dasyra} \& {Dumke}}{{Alton} et~al.}{2004}]{2004A&A...425..109A} {Alton}, P.~B., {Xilouris}, E.~M., {Misiriotis}, A., {Dasyra}, K.~M., {Dumke}, M.\ 2004, \aap, 425, 109

\bibitem[\protect\citeauthoryear{{Aniano}, {Draine}, {Gordon} \& {Sandstrom}}{{Aniano} et~al.}{2011}]{2011PASP..123.1218A} {Aniano}, G., {Draine}, B.~T., {Gordon}, K.~D., {Sandstrom}, K.\  2011, \pasp, 123, 1218

\bibitem[\protect\citeauthoryear{{Bahcall} \& {Soneira}}{{Bahcall} \& {Soneira}}{1980}]{1980ApJS...44...73B} {Bahcall}, J.~N., {Soneira}, R.~M.\ 1980, \apjs, 44, 73


\bibitem[Baes et al.(2003)]{2003MNRAS.343.1081B} Baes, M., Davies, J.~I., Dejonghe, H., et al.\ 2003, \mnras, 343, 1081 

\bibitem[Baes et al.(2010)]{2010A&A...518L..39B} Baes, M., Fritz, J., Gadotti, D.~A., et al.\ 2010, \aap, 518, L39   
  
\bibitem[Baes \& Gentile(2011)]{2011A&A...525A.136B} Baes, M., \& Gentile, G.\ 2011, \aap, 525, A136 

\bibitem[Baes \& Van Hese(2011)]{2011A&A...534A..69B} Baes, M., \& Van Hese, E.\ 2011, \aap, 534, A69

\bibitem[\protect\citeauthoryear{{Baes}, {Verstappen}, {De Looze}, {Fritz}, {Saftly}, {Vidal P{\'e}rez}, {Stalevski} \& {Valcke}}{{Baes} et~al.}{2011}]{2011ApJS..196...22B} {Baes}, M., {Verstappen}, J., {De Looze}, I., {Fritz}, J., {Saftly}, W., {Vidal P{\'e}rez}, E., {Stalevski}, M., {Valcke}, S.\  2011, \apjs, 196, 22  

\bibitem[Baes \& Viaene(2016)]{2016A&A...587A..86B} Baes, M., \& Viaene, S.\ 2016, \aap, 587, A86  

\bibitem[Barteldrees \& Dettmar(1994)]{1994A&AS..103..475B} Barteldrees, A., \& Dettmar, R.-J.\ 1994, \aaps, 103,  

\bibitem[Bianchi et al.(2000)]{2000A&A...359...65B} Bianchi, S., Davies, J.~I., \& Alton, P.~B.\ 2000, \aap, 359, 65  
  
\bibitem[\protect\citeauthoryear{{Bianchi}}{{Bianchi}}{2007}]{2007A&A...471..765B} {Bianchi}, S.\ 2007, \aap, 471, 765

\bibitem[\protect\citeauthoryear{{Bianchi}}{{Bianchi}}{2008}]{2008A&A...490..461B} {Bianchi}, S.\ 2008, \aap, 490, 461

\bibitem[Bianchi \& Xilouris(2011)]{2011A&A.531L.11B} Bianchi, S., \& Xilouris, E.~M.\ 2011, \aap, 531, L11 

\bibitem[Bianchi et al.(2014)]{2014AdSpR..53..900B} Bianchi, L., Conti, A., \& Shiao, B.\ 2014, Advances in Space Research, 53, 900 

\bibitem[Bizyaev \& Mitronova(2009)]{2009ApJ...702.1567B} Bizyaev, D., \& Mitronova, S.\ 2009, \apj, 702, 1567 

\bibitem[Bobylev \& Bajkova(2016)]{2016AstL...42....1B} Bobylev, V.~V., \& Bajkova, A.~T.\ 2016, Astronomy Letters, 42, 1 

\bibitem[\protect\citeauthoryear{{Bruzual} \& {Charlot}}{{Bruzual} \& {Charlot}}{2003}]{2003MNRAS.344.1000B} {Bruzual}, G., {Charlot}, S.\ 2003, \mnras, 344, 1000

\bibitem[Bureau et al.(2006)]{2006MNRAS.370..753B} Bureau, M., Aronica, G., Athanassoula, E., et al.\ 2006, \mnras, 370, 753 

\bibitem[\protect\citeauthoryear{{Camps} \& {Baes}}{{Camps} \& {Baes}}{2015}]{2015A&C.....9...20C} {Camps}, P., {Baes}, M.\ 2015, Astronomy and Computing, 9, 20

\bibitem[Camps et al.(2015)]{2015A&A...580A..87C} Camps, P., Misselt, K., Bianchi, S., et al.\ 2015, \aap, 580, A87 

\bibitem[\protect\citeauthoryear{Caon, Capaccioli \& D'Onofrio}{1993}]{caon+1993} Caon, N., Capaccioli, M., D'Onofrio, M.\ 1993, \mnras, 265, 1013

\bibitem[Cardelli et al.(1989)]{1989ApJ...345..245C} Cardelli, J.~A., Clayton, G.~C., \& Mathis, J.~S.\ 1989, \apj, 345, 245 

\bibitem[\protect\citeauthoryear{{Chabrier}}{{Chabrier}}{2003}]{2003PASP..115..763C}{Chabrier}, G.\ 2003, \pasp, 115, 763

\bibitem[Chiba \& Beers(2000)]{2000AJ....119.2843C} Chiba, M., \& Beers, T.~C.\ 2000, \aj, 119, 2843 

\bibitem[{{Ciotti} \& {Bertin}(1999)}]{ciotti99}{Ciotti}, L. \& {Bertin}, G. 1999, A\&A, 352, 447

\bibitem[Comer{\'o}n et al.(2011)]{2011ApJ...741...28C} Comer{\'o}n, S., Elmegreen, B.~G., Knapen, J.~H., et al.\ 2011, \apj, 741, 28 

\bibitem[Comer{\'o}n et al.(2012)]{2012ApJ...759...98C} Comer{\'o}n, S., Elmegreen, B.~G., Salo, H., et al.\ 2012, \apj, 759, 98 

\bibitem[Compi{\`e}gne et al.(2011)]{2011A&A...525A.103C} Compi{\`e}gne, M., Verstraete, L., Jones, A., et al.\ 2011, \aap, 525, A103 

\bibitem[Corradi et al.(1996)]{1996ASSL..209..523C} Corradi, R.~L.~M., Beckman, J.~E., del Rio, M.~S., di Bartolomeo, A., \& Simonneau, E.\ 1996, New Extragalactic Perspectives in the New South Africa, 209, 523 

\bibitem[\protect\citeauthoryear{{Dasyra}, {Xilouris}, {Misiriotis} \& {Kylafis}}{{Dasyra} et~al.}{2005}]{2005A&A...437..447D}{Dasyra}, K.~M., {Xilouris}, E.~M., {Misiriotis}, A., {Kylafis}, N.~D.\ 2005, \aap, 437, 447

\bibitem[\protect\citeauthoryear{{De Geyter}, {Baes}, {Fritz} \& {Camps}}{{De Geyter} et~al.}{2013}]{2013A&A...550A..74D}{De Geyter}, G., {Baes}, M., {Fritz}, J., {Camps}, P.\ 2013, \aap, 550, A74

\bibitem[\protect\citeauthoryear{{De Geyter}, {Baes}, {Camps}, {Fritz}, {De Looze}, {Hughes}, {Viaene} \& {Gentile}}{{De Geyter} et~al.}{2014}]{2014MNRAS.441..869D} {De Geyter}, G., {Baes}, M., {Camps}, P., {Fritz}, J., {De Looze}, I., {Hughes}, T.~M., {Viaene}, S., {Gentile}, G.\ 2014, \mnras, 441, 869

\bibitem[De Geyter et al.(2015)]{2015MNRAS.451.1728D} De Geyter, G., Baes, M., De Looze, I., et al.\ 2015, \mnras, 451, 1728 (DG15)

\bibitem[\protect\citeauthoryear{{De Looze}, {Baes}, {Fritz} \& {Verstappen}}{{De Looze} et~al.}{2012a}]{2012MNRAS.419..895D}{De Looze}, I., {Baes}, M., {Fritz}, J., {Verstappen}, J.\ 2012a, \mnras, 419, 895



\bibitem[De Looze et al.(2014)]{2014A&A...571A..69D} De Looze, I., Fritz, J., Baes, M., et al.\ 2014, \aap, 571, A69 

\bibitem[Draine \& Li(2001)]{2001ApJ...551..807D} Draine, B.~T., \& Li, A.\ 2001, \apj, 551, 807

\bibitem[Draine \& Li(2007)]{2007ApJ...657..810D} Draine, B.~T., \& Li, A.\ 2007, \apj, 657, 810
  
\bibitem[Eales et al.(2010)]{2010PASP..122..499E} Eales, S., Dunne, L., Clements, D., et al.\ 2010, \pasp, 122, 499 

\bibitem[{Erwin {et~al.}(2005)Erwin, Beckman, \& Pohlen}]{erwin05} Erwin, P., Beckman, J.~E., \& Pohlen, M. 2005, ApJL, 626, L81

\bibitem[{{Erwin} {et~al.}(2008){Erwin}, {Pohlen}, \& {Beckman}}]{erwin08} {Erwin}, P., {Pohlen}, M., \& {Beckman}, J.~E. 2008, AJ, 135, 20

\bibitem[Erwin(2015)]{2015ApJ...799..226E} Erwin, P.\ 2015, \apj, 799, 226 

\bibitem[Eskew et al.(2012)]{2012AJ....143..139E} Eskew, M., Zaritsky, D., \& Meidt, S.\ 2012, \aj, 143, 139   
  
\bibitem[Fabricius et al.(2012)]{2012ApJ...754...67F} Fabricius, M.~H., Saglia, R.~P., Fisher, D.~B., et al.\ 2012, \apj, 754, 67 

\bibitem[Fanciullo et al.(2015)]{2015A&A...580A.136F} Fanciullo, L., Guillet, V., Aniano, G., et al.\ 2015, \aap, 580, A136

\bibitem[Fazio et al.(2004)]{2004ApJS..154...10F} Fazio, G.~G., Hora, J.~L., Allen, L.~E., et al.\ 2004, \apjs, 154, 10 

\bibitem[Fisher \& Drory(2010)]{2010ApJ...716..942F} Fisher, D.~B., \& Drory, N.\ 2010, \apj, 716, 942 


\bibitem[Gilmore \& Reid(1983)]{1983MNRAS.202.1025G} Gilmore, G., \& Reid, N.\ 1983, \mnras, 202, 1025 

\bibitem[Goldsmith et al.(2007)]{2007ApJ.654.273G} Goldsmith, P.~F., Li, D., \& Kr{\v c}o, M.\ 2007, \apj, 654, 273 

\bibitem[Gordon et al.(2003)]{2003ApJ...594..279G} Gordon, K.~D., Clayton, G.~C., Misselt, K.~A., Landolt, A.~U., \& Wolff, M.~J.\ 2003, \apj, 594, 279 

\bibitem[\protect\citeauthoryear{{Gordon}, {Misselt}, {Witt} \& {Clayton}}{{Gordon} et~al.}{2001}]{2001ApJ...551..269G} {Gordon}, K.~D., {Misselt}, K.~A., {Witt}, A.~N., {Clayton}, G.~C.\ 2001, \apj, 551, 269

\bibitem[Griffin et al.(2010)]{2010A&A...518L...3G} Griffin, M.~J., Abergel, A., Abreu, A., et al.\ 2010, \aap, 518, L3   

\bibitem[\protect\citeauthoryear{{Guhathakurta} \& {Draine}}{{Guhathakurta} \& {Draine}}{1989}]{1989ApJ...345..230G} {Guhathakurta}, P., {Draine}, B.~T.\ 1989, \apj, 345, 230

\bibitem[Hatziminaoglou et al.(2008)]{2008MNRAS.386.1252H} Hatziminaoglou, E., Fritz, J., Franceschini, A., et al.\ 2008, \mnras, 386, 1252   

\bibitem[\protect\citeauthoryear{{Hatziminaoglou}, {Fritz} \& {Jarrett}}{{Hatziminaoglou} et~al.}{2009}]{2009MNRAS.399.1206H} {Hatziminaoglou}, E., {Fritz}, J., {Jarrett}, T.~H.\ 2009, \mnras, 399, 1206

\bibitem[Hodges-Kluck \& Bregman(2014)]{2014ApJ...789..131H} Hodges-Kluck, E., \& Bregman, J.~N.\ 2014, \apj, 789, 131 

\bibitem[Holwerda et al.(2012)]{2012A&A...541L...5H} Holwerda, B.~W., Bianchi, S., B{\"o}ker, T., et al.\ 2012, \aap, 541, L5   

\bibitem[Hughes et al.(2014)]{2014A&A...565A...4H} Hughes, T.~M., Baes, M., Fritz, J., et al.\ 2014, \aap, 565, A4 

\bibitem[Hughes et al.(2015)]{2015A&A.575A.17H} Hughes, T.~M., Foyle, K., Schirm, M.~R.~P., et al.\ 2015, \aap, 575, A17 

\bibitem[Jones et al.(2013)]{2013A&A...558A..62J} Jones, A.~P., Fanciullo, L., K{\"o}hler, M., et al.\ 2013, \aap, 558, A62   
  
\bibitem[\protect\citeauthoryear{{Jonsson}}{{Jonsson}}{2006}]{2006MNRAS.372....2J}{Jonsson}, P.\ 2006, \mnras, 372, 2

\bibitem[Juri{\'c} et al.(2008)]{2008ApJ...673..864J} Juri{\'c}, M., Ivezi{\'c}, {\v Z}., Brooks, A., et al.\ 2008, \apj, 673, 864 

\bibitem[K{\"o}hler et al.(2014)]{2014A&A...565L...9K} K{\"o}hler, M., Jones, A., \& Ysard, N.\ 2014, \aap, 565, L9 

\bibitem[K{\"o}hler et al.(2015)]{2015A&A...579A..15K} K{\"o}hler, M., Ysard, N., \& Jones, A.~P.\ 2015, \aap, 579, A15 

\bibitem[Kregel et al.(2004)]{2004MNRAS.352..768K} Kregel, M., van der Kruit, P.~C., \& de Blok, W.~J.~G.\ 2004, \mnras, 352, 768 

\bibitem[\protect\citeauthoryear{Kregel, van der Kruit \& de Grijs}{2002}]{kregel+2002} Kregel, M., van~der~Kruit, P.C., de~Grijs, R.\ 2002, \mnras, 334, 646

\bibitem[Kylafis \& Bahcall(1987)]{1987ApJ...317..637K} Kylafis, N.~D., \& Bahcall, J.~N.\ 1987, \apj, 317, 637 

\bibitem[Larsen \& Humphreys(2003)]{2003AJ....125.1958L} Larsen, J.~A., \& Humphreys, R.~M.\ 2003, \aj, 125, 1958 

\bibitem[Leitherer et al.(1999)]{1999ApJS..123....3L} Leitherer, C., Schaerer, D., Goldader, J.~D., et al.\ 1999, \apjs, 123, 3   
  

\bibitem[Martin et al.(2005)]{2005ApJ...619L...1M} Martin, D.~C., Fanson, J., Schiminovich, D., et al.\ 2005, \apjl, 619, L1 

\bibitem[Mart{\'{\i}}n-Navarro et al.(2012)]{2012MNRAS.427.1102M} Mart{\'{\i}}n-Navarro, I., Bakos, J., Trujillo, I., et al.\ 2012, \mnras, 427, 1102 

\bibitem[Mazure \& Capelato(2002)]{2002A&A...383..384M} Mazure, A., \& Capelato, H.~V.\ 2002, \aap, 383, 384 

\bibitem[McMillan(2011)]{2011MNRAS.414.2446M} McMillan, P.~J.\ 2011, \mnras, 414, 2446 

\bibitem[Metcalfe et al.(2000)]{2000ApJ...545..974M} Metcalfe, T.~S., Nather, R.~E., \& Winget, D.~E.\ 2000, \apj, 545, 974 

\bibitem[Misiriotis et al.(2001)]{2001A&A...372..775M} Misiriotis, A., Popescu, C.~C., Tuffs, R., \& Kylafis, N.~D.\ 2001, \aap, 372, 775 

\bibitem[Mosenkov et al.(2010)]{2010MNRAS.401.559M} Mosenkov, A.~V., Sotnikova, N.~Y., \& Reshetnikov, V.~P.\ 2010, \mnras, 401, 559

\bibitem[Mosenkov et al.(2015)]{2015MNRAS.451.2376M} Mosenkov, A.~V., Sotnikova, N.~Y., Reshetnikov, V.~P., Bizyaev, D.~V., \& Kautsch, S.~J.\ 2015, \mnras, 451, 2376  

\bibitem[{Mu{\~n}oz-Mateos {et~al.}(2013)Mu{\~n}oz-Mateos, Sheth, Gil~de Paz, Meidt, Athanassoula, Bosma, {Comer{\'o}n}, {Elmegreen}, \& {Elmegreen}}]{munoz-mateos13} Mu{\~n}oz-Mateos, J.~C., Sheth, K., Gil~de Paz, A., Meidt, S., Athanassoula, E., Bosma, A., {Comer{\'o}n}, S., {Elmegreen}, D.~M., \& {Elmegreen}, B.~G.\ 2013, ApJ, 771, 59

\bibitem[O'Brien et al.(2010a)]{2010A&A...515A..60O} O'Brien, J.~C., Freeman, K.~C., van der Kruit, P.~C., \& Bosma, A.\ 2010, \aap, 515, A60 

\bibitem[O'Brien et al.(2010b)]{2010A&A...515A..62O} O'Brien, J.~C., Freeman, K.~C., \& van der Kruit, P.~C.\ 2010, \aap, 515, A62 

\bibitem[Patsis \& Xilouris(2006)]{2006MNRAS.366.1121P} Patsis, P.~A., \& Xilouris, E.~M.\ 2006, \mnras, 366, 1121 

\bibitem[Peletier et al.(1994)]{1994A&AS..108..621P} Peletier, R.~F., Valentijn, E.~A., Moorwood, A.~F.~M., \& Freudling, W.\ 1994, \aaps, 108,  

\bibitem[Pilbratt et al.(2010)]{2010A&A...518L...1P} Pilbratt, G.~L., Riedinger, J.~R., Passvogel, T., et al.\ 2010, \aap, 518, L1 

\bibitem[Planck Collaboration et al.(2014)]{2014A&A...571A...1P} Planck Collaboration, Ade, P.~A.~R., Aghanim, N., et al.\ 2014, \aap, 571, A1 

\bibitem[\protect\citeauthoryear{{Poglitsch} et~al.,}{{Poglitsch} et~al.}{2010}]{2010A&A...518L...2P} {Poglitsch}, A. et~al.\ 2010, \aap, 518, L2

\bibitem[\protect\citeauthoryear{Pohlen et al.}{2000}]{pohlen+2000} Pohlen, M., Dettmar, R.-J., L\.{u}tticke, R., Schwarzkopf, U.\ 2000, \aaps, 144, 405

\bibitem[Popescu et al.(2000)]{2000A&A...362..138P} Popescu, C.~C., Misiriotis, A., Kylafis, N.~D., Tuffs, R.~J., \& Fischera, J.\ 2000, \aap, 362, 138 

\bibitem[Popescu \& Tuffs(2002)]{2002MNRAS.335L..41P} Popescu, C.~C., \& Tuffs, R.~J.\ 2002, \mnras, 335, L41 

\bibitem[Popescu et al.(2011)]{2011A&A...527A.109P} Popescu, C.~C., Tuffs, R.~J., Dopita, M.~A., et al.\ 2011, \aap, 527, A109 

\bibitem[Reid \& Majewski(1993)]{1993ApJ...409..635R} Reid, N., \& Majewski, S.~R.\ 1993, \apj, 409, 635 

\bibitem[\protect\citeauthoryear{{Robitaille}}{{Robitaille}}{2011}]{2011A&A...536A..79R}{Robitaille}, T.~P.\ 2011, \aap, 536, A79

\bibitem[Saftly et al.(2015)]{2015A&A...576A..31S} Saftly, W., Baes, M., De Geyter, G., et al.\ 2015, \aap, 576, A31 

\bibitem[Salo et al.(2015)]{2015ApJS..219....4S} Salo, H., Laurikainen, E., Laine, J., et al.\ 2015, \apjs, 219, 4 

\bibitem[\protect\citeauthoryear{{Salpeter}}{{Salpeter}}{1955}]{1955ApJ...121..161S}{Salpeter}, E.~E.\ 1955, \apj, 121, 161

\bibitem[Sandin(2014)]{2014A&A...567A..97S} Sandin, C.\ 2014, \aap, 567, A97 

\bibitem[Schechtman-Rook et al.(2012)]{2012ApJ...746...70S} Schechtman-Rook, A., Bershady, M.~A., \& Wood, K.\ 2012, \apj, 746, 70 

\bibitem[\protect\citeauthoryear{{Schechtman-Rook} \& {Bershady}}{{Schechtman-Rook} \& {Bershady}}{2013}]{2013ApJ...773...45S}{Schechtman-Rook}, A., {Bershady}, M.~A.\  2013, \apj, 773, 45

\bibitem[\protect\citeauthoryear{{Schechtman-Rook} \& {Bershady}}{{Schechtman-Rook} \& {Bershady}}{2014}]{2014ApJ...795..136S} {Schechtman-Rook}, A., {Bershady}, M.~A.\  2014, \apj, 795, 136

\bibitem[Schlafly \& Finkbeiner(2011)]{2011ApJ...737..103S} Schlafly, E.~F., \& Finkbeiner, D.~P.\ 2011, \apj, 737, 103 

\bibitem[\protect\citeauthoryear{S\'ersic}{1968}]{ser1968} S\'ersic, J.L.\ 1968, Atlas de Galaxias Australes, Observatorio Astronomico, Cordoba


\bibitem[Shinn \& Seon(2015)]{2015ApJ...815..133S} Shinn, J.-H., \& Seon, K.-I.\ 2015, \apj, 815, 133 

\bibitem[Skrutskie et al.(2006)]{2006AJ....131.1163S} Skrutskie, M.~F., Cutri, R.~M., Stiening, R., et al.\ 2006, \aj, 131, 1163 

\bibitem[Soifer \& Neugebauer(1991)]{1991AJ.101.354S} Soifer, B.~T., \& Neugebauer, G.\ 1991, \aj, 101, 354 

\bibitem[Steinacker et al.(2013)]{2013ARA&A..51...63S} Steinacker, J., Baes, M., \& Gordon, K.~D.\ 2013, \araa, 51, 63 

\bibitem[Tielens(2005)]{Tielens2005} Tielens, A.~G.~G.~M.\ 2005, The Physics and Chemistry of the Interstellar Medium, Cambridge University Press

\bibitem[\protect\citeauthoryear{van~der~Kruit \& Searle}{1981a}]{vderk1981a} van~der~Kruit~P.C., Searle~L.\ 1981a, \aap, 95, 105

\bibitem[\protect\citeauthoryear{van~der~Kruit \& Searle}{1981b}]{vderk1981b} van~der~Kruit, P.C., Searle, L.\ 1981b, \aap, 95, 116


\bibitem[Verstappen et al.(2013)]{2013A&A...556A..54V} Verstappen, J., Fritz, J., Baes, M., et al.\ 2013, \aap, 556, A54 

\bibitem[Viaene et al.(2015)]{2015A&A...579A.103V} Viaene, S., De Geyter, G., Baes, M., et al.\ 2015, \aap, 579, A103 

\bibitem[Viaene et al.(2016)]{2016A&A...586A..13V} Viaene, S., Baes, M., Bendo, G., et al.\ 2016, \aap, 586, A13 


\bibitem[Werner et al.(2004)]{2004ApJS..154....1W} Werner, M.~W., Roellig, T.~L., Low, F.~J., et al.\ 2004, \apjs, 154, 1 

\bibitem[Wright et al.(2010)]{2010AJ....140.1868W} Wright, E.~L., Eisenhardt, P.~R.~M., Mainzer, A.~K., et al.\ 2010, \aj, 140, 1868-1881 

\bibitem[Xilouris et al.(1997)]{1997A&A...325..135X} Xilouris, E.~M., Kylafis, N.~D., Papamastorakis, J., Paleologou, E.~V., \& Haerendel, G.\ 1997, \aap, 325, 135 

\bibitem[Xilouris et al.(1998)]{1998A&A.331.894X} Xilouris, E.~M., Alton, P.~B., Davies, J.~I., et al.\ 1998, \aap, 331, 894 

\bibitem[Xilouris et~al.(1999)]{1999A&A.344.868X} Xilouris, E.~M., Byun, Y.~I., Kylafis, N.~D., Paleologou, E.~V., Papamastorakis, J.\ 1999, \aap, 344, 868 (X99)

\bibitem[Zubko et al.(2004)]{2004ApJS..152..211Z} Zubko, V., Dwek, E., \& Arendt, R.~G.\ 2004, \apjs, 152, 211 

\end{thebibliography}

\end{document}